\documentclass[pre, amsmath, amssymb, superscriptaddress]{revtex4}

\usepackage[utf8]{inputenc}
\usepackage{graphicx}
\usepackage[colorlinks]{hyperref}

\usepackage{amsmath}

\usepackage{graphicx}
\usepackage{dcolumn}
\usepackage{bm}
\usepackage{braket}
\usepackage{SIunits} 

\newcommand{\beq}{\begin{equation}}
\newcommand{\eeq}{\end{equation}}

\begin{document}

\title{Shape Complementarity Optimization of Antibody-Antigen Interfaces: the Application to SARS-CoV-2 Spike Protein}

\author{Alfredo De Lauro \footnote[2]{ These authors contributes equally to this work.}}
\affiliation{Department of Sciences, Roma Tre University, 00146 Rome, Italy}
\author{Lorenzo Di Rienzo \footnotemark[2],\footnote{Corresponding author: \\ lorenzo.dirienzo@iit.it} }
\affiliation{Center for Life Nano \& Neuro-Science, Istituto Italiano di Tecnologia, Viale Regina Elena 291, 00161, Rome, Italy}
\author{Mattia Miotto}
\affiliation{Center for Life Nano \& Neuro-Science, Istituto Italiano di Tecnologia, Viale Regina Elena 291, 00161, Rome, Italy}
\author{Pier Paolo Olimpieri}
\affiliation{Department of Physics, Sapienza University, Piazzale Aldo Moro 5, 00185, Rome, Italy}
\author{Edoardo Milanetti}
\affiliation{Department of Physics, Sapienza University, Piazzale Aldo Moro 5, 00185, Rome, Italy}
\affiliation{Center for Life Nano \& Neuro-Science, Istituto Italiano di Tecnologia, Viale Regina Elena 291, 00161, Rome, Italy}
\author{Giancarlo Ruocco}
\affiliation{Center for Life Nano \& Neuro-Science, Istituto Italiano di Tecnologia, Viale Regina Elena 291, 00161, Rome, Italy}
\affiliation{Department of Physics, Sapienza University, Piazzale Aldo Moro 5, 00185, Rome, Italy}

\begin{abstract}
Many factors influence biomolecules binding, and its assessment constitutes an elusive challenge in computational structural biology. In this respect, the evaluation of shape complementarity at molecular interfaces is one of the main factors to be considered. We focus on the particular case of antibody-antigen complexes to quantify the complementarities occurring at molecular interfaces. We relied on a method we recently developed, which employs the 2D Zernike descriptors, to characterize investigated regions with an ordered set of numbers summarizing the local shape properties. Collected a structural dataset of antibody-antigen complexes,  we applied this method and we statistically distinguished, in terms of shape complementarity, pairs of interacting regions from non-interacting ones. Thus, we set up a novel computational strategy based on \textit{in-silico} mutagenesis of antibody binding site residues. We developed a Monte Carlo procedure to increase the shape complementarity between the antibody paratope and a given epitope on a target protein surface. We applied our protocol against several molecular targets in SARS-CoV-2 spike protein, known to be indispensable for viral cell invasion. We, therefore, optimized the shape of template antibodies for the interaction with such regions. As the last step of our procedure, we performed an independent molecular docking validation of the results of our Monte Carlo simulations.

\end{abstract}

\maketitle

\section{Introduction}

Cellular functioning is widely dependent on processes occurring when biological molecules recognize each other and bind \cite{jones1996principles,gromiha2017protein}. In particular, the non-covalent protein-protein pairing proved to be essential in several biochemical pathways, ranging from biocatalysis to organism immunity or cell regulatory network construction \cite{gavin2002functional,han2004evidence}.
Not surprisingly, in the last decades, a very high effort has been devoted to developing computational tools for the structural characterization of protein-protein complexes. The aim of these methods are various, varying from binding site identification \cite{gainza2020deciphering, milanetti20212d} to binding affinity prediction \cite{siebenmorgen2020computational, vangone2015contacts} or protein-protein docking guide \cite{vakser2014protein, kozakov2017cluspro, geng2020iscore, yan2020hdock}.

In this scenario, the shape complementarity at the molecular interface is one of the most basic tasks to take into account \cite{jones1996principles, katchalski1992molecular, lawrence1993shape}.
 Indeed, the evaluation of shape complementarity is essential for docking, both in terms of searching and evaluating the binding poses \cite{chen2003novel,nicola2007simple, kuroda2016shape, gromiha2017protein, yan2019pushing}, and represents one of the factors to take into account for binding site recognition \cite{gainza2020deciphering, milanetti20212d} or to assess the binding affinity \cite{erijman2014structure}.
 
 Among the wide variety of methods developed in the last years to describe the geometrical properties of a molecular region and to evaluate the complementarity with a putative binding partner region, using the Zernike polynomials is an effective and promising strategy \cite{venkatraman2009potential, di2017superposition, di2021quantitative, daberdaku2019antibody}. Indeed, once extracted the molecular surface region, its geometrical properties are summarized through a set of numerical descriptors, namely the Zernike descriptors. The accuracy of the description is increased by enlarging the number of descriptors considered \cite{zernike1934diffraction, novotni2004shape, canterakis19993d}.

The main advantage of the Zernike formalism is that the molecular surface representation is invariant under protein rotation, constituting an absolute morphological characterization of the examined protein region. Therefore, the complementarity between two molecular regions is computed by comparing their Zernike descriptors, without the need for any preliminary superposition step\cite{daberdaku2018exploring, di2020quantitative}.

In the last decade, the Zernike approach, in its 3D version, has been widely applied for the analysis of biomolecules \cite{di2017superposition,venkatraman2009protein, kihara2011molecular, alba2020molecular, daberdaku2019antibody, di2020quantitative, venkatraman2009potential, daberdaku2018exploring, han2019global,di2020novel}, proving its efficacy in characterizing both global and local proteins properties.

We recently developed a computational protocol that allows us to employ the 2D Zernike formalism to assess the shape complementarity observed in protein-protein interfaces \cite{milanetti20212d}. The utilization of the 2D formalism allows to sensibly decrease the computational time needed to compute the shape descriptors without significant loss in description accuracy \cite{di2021computational}.

In this work, we focused on Antibody-Antigen interactions, since these complexes represent a critical case of molecular recognition where interface shape complementarity level is similar to the typical protein-protein interfaces \cite{kuroda2016shape,li2003x}.

Moreover, antibodies have been the object of extensive biomedical studies since their modular architecture facilitates the engineering of novel binding sites \cite{singh2018monoclonal, saeed2017antibody, gotwals2017prospects}. Indeed, the recognition of virtually any foreign antigen is due to high sequence variability in the antigen-binding site, while the overall architecture is largely conserved \cite{chothia1987canonical, chothia1989conformations, tramontano1990framework}. The antigen-binding site is structurally composed of three loops of both the heavy and light chains, forming the \textit{Complementary Determining Regions} (CDR). Notwithstanding the variability of the CDR sequences, these loops (at least five out of six) can acquire only a limited number of structural conformations, called \textit{canonical structure}. Moreover, studying the growing number of experimentally determined antibody structures, it has been demonstrated the relationship between the presence of given residues in certain sequence positions and the canonical structure adopted by the antibody \cite{tramontano1990framework, chothia1992structural,foote1992antibody, decanniere2000canonical, chailyan2011structural, north2011new}.

In this framework, thanks to the public availability of an increasing number of experimental antibodies structures \cite{dunbar2013sabdab}, several very effective computational approaches -for predicting the structure of antibodies from their sequences- have been produced, often based on machine learning approaches \cite{dunbar2016sabpred,lepore2017pigspro,weitzner2017modeling, Abanades2022ABlooper}. Moreover, obtaining structural information about Antibody-Antigen complexes has been the object of extensive studies and it is still elusive. Many computational protocols focused on the prediction of the residues involved in partner interaction, both on the antibody and antigen side of the interface \cite{olimpieri2013prediction,liberis2018parapred, potocnakova2016introduction}. 

All these kinds of computational tools can be used for antibody design, that is the development of a novel molecule able to bind a given antigen \cite{norman2020computational}. In particular, \textit{ab initio} protocols are able to design paratopes integrating antibody structure prediction, molecular docking and binding energy assessment \cite{ pantazes2010optcdr, li2014optmaven,adolf2018rosettaantibodydesign,lapidoth2015abdesign}.

Here, we collected a structural dataset of antibody-protein complexes solved in x-ray crystallography. In this work, we apply for the first time our recently developed method based on 2D Zernike descriptors to study the antibody-antigen interfaces. Concerning this specific kind of interaction, we demonstrate that such a fast and compact description can recognize with satisfying success the specific interaction from non-specific ones. Indeed, paratopes show a shape complementarity statistically higher toward their corresponding epitopes than toward epitopes belonging to unrelated antigens. 

Based on these results, we propose here for the first time a new computational protocol employing 2D Zernike descriptors that, for a given target protein region, optimizes the shape complementarity of an antibody toward that region. 
Indeed, once a target region, belonging to a protein antigen, is identified and characterized with its Zernike descriptors, we compared such region with the paratope of the antibodies in our dataset. Selecting as starting template the antibody that has the most complementary patch, we perform a Monte Carlo (MC) simulation for the optimization of the paratope structural conformation. Through extensive computational mutagenesis, substituting in each step an interacting antibody residue with a different random one, we accept or reject each mutation according to the gain in shape complementarity, as evaluated by to Zernike method \cite{di2021computational, di2020novel}. In this work, the combined application of both the 2D Zernike formalism and a Monte Carlo simulation allows a computationally fast and effective exploration of the space of the possible mutants.

In the current pandemic situation, the interactions between SARS-CoV-2 spike protein and human cellular receptors have been extensively studied through 2D Zernike polynomials formalism \cite{milanetti2020silico, Miotto2021Molecular, bo2021exploring, miotto2022inferring}. Therefore, despite the generality of such an approach, we selected as a target for the optimization protocol some surface regions of SARS-CoV-2 spike protein. We discuss here the results we got. Indeed, elucidating the interaction mechanism between antibodies and viral proteins represents a fundamental element for developing new therapies.

\section{Results and Discussion}

\subsection{Description of Antibody-Antigen interface through Zernike Descriptors}

\begin{figure*}[!]
\centering
\includegraphics[width =0.8 \textwidth]{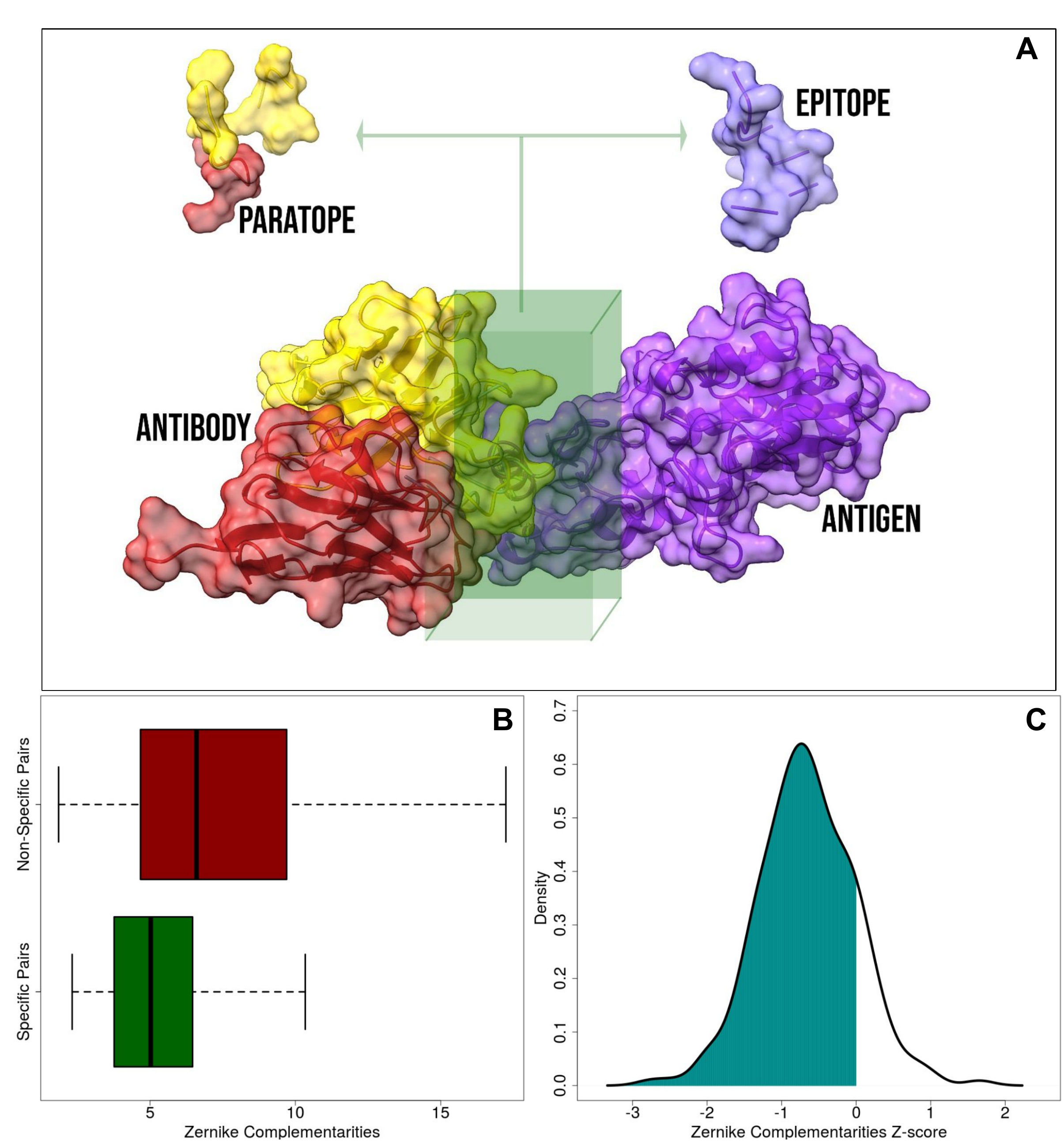}
\caption{\textbf{Application of the 2D Zernike polynomials approach to antibody-antigen complexes.}
\textbf{A)} Representation of a molecular antibody-antigen complex: the antibody heavy chain, antibody light chain, and the antigen are in red, yellow, and purple, respectively. The interacting regions, defined as the portion of the molecular surfaces belonging to residues closer than 4 $\AA$ to any atoms of the molecular partner, are extracted from the whole surface. \textbf{B)} Boxplot comparing the \textit{specific complementarity}, i.e. the complementarity between regions actually found in interaction (green), and the \textit{non-specific complementarity}, i.e. the complementarity observed between paratopes and epitopes of different complexes (red). It is worth remarking that when the numerical value is low, the complementarity is high. \textbf{C)} Z-scores distribution of the specific complementarity. When the Z-score is lower than 0, the specific interaction is characterized by a complementarity higher than the mean of the non-specific interaction. 
}
\label{fig:metodo_dataset}
\end{figure*}

In the present section, we discuss the results we obtained applying our recently developed computational protocol \cite{milanetti20212d} on a structural dataset composed of 229 antibody-antigen complexes (See Methods).

In particular, we firstly identified for each complex the paratope (epitope) as the set of residues with at least one atom closer than 4 $\AA$ to an antigen (antibody) atom. Therefore, after separately computing the molecular surface \cite{richards1977areas} for both the proteins in interaction, we extracted the portions of the molecular surfaces relating to the binding site residues to properly characterize the shape of the interacting regions of antibodies and antigens ( See Figure \ref{fig:metodo_dataset}.A). 

Once identified the interacting regions, we characterized them through the 2D Zernike polynomials, summarizing their geometrical properties in an ordered set of numerical descriptors (See Materials and Methods). By definition, two perfectly fitting surfaces are characterized by the same shape,  meaning that, in principle, the difference between their Zernike descriptors is zero. Therefore, the shape complementarity between two molecular regions is compactly evaluated by such formalism. The lower is the distance between the Zernike descriptors, the higher is the shape complementarity between the corresponding protein regions \cite{milanetti20212d, di2021computational}.

We described all the paratopes and epitopes in the dataset with the Zernike formalism. In summary, we deal with 229 (the number of structures in our dataset) sets of 121 ( the number of invariant descriptors when the order of expansion is set to 20) numerical descriptors for the paratopes and 229 for the epitopes. We thus defined the \textit{specific complementarities} as the euclidean distance between the descriptors of the 229 pairs of interacting (extracted from the same structure) paratope and epitope. Diversely, \textit{Non-Specific complementarity} is the euclidean distance between all the pairs of unrelated paratopes and epitopes (i.e. a paratope extracted from one structure and an epitope extracted from another one). In the end, we deal with 229 specific complementarities (one for each complex) and a high number (~25000) of non-specific complementarities (all the possible paratope-epitope pairs given a dataset of 229 items). In other words, in dealing with $N$ antibody-antigen complexes, the specific complementarity, $C_s$,  is defined as:

\begin{equation}
    D(p_{i}, e_{j})|_{i=j} = \sqrt{\sum_{k=1}^{121} (p_i^k - e_j^k)^2} |_{i=j}
    \label{eq:spec_dist}
\end{equation}

where $D$ is the Euclidean distance between the vectors of the paratope ($p_i$) and epitope ($e_i$) Zernike descriptors. Since we expanded all the paratopes and epitopes to order 20, we dealt with 121 descriptors for each binding region. On the other hand, the non-specific complementarities, $C_{ns},$ can be computed as:

\begin{equation}
    D(p_{i}, e_{j})|_{i \neq j}  = \sqrt{\sum_{k=1}^{121} (p_i^k - e_j^k)^2} |_{i \neq j}
    \label{eq:nonspec_dist}
\end{equation}

In Figure \ref{fig:metodo_dataset}.B we reported a boxplot highlighting the differences between $C_{s}$ and $C_{ns}$ distributions. As expected, the distribution of $C_s$ is statistically lower than the distribution of $C_{ns}$ (Kolmogorov-Smirnov test p-value $< 2.2e-16$), testifying the sensitivity of Zernike in recognizing regions actually in contact from non-interacting ones. 
In the second step, we normalized the Zernike complementarities with the Z-score. In particular, for each paratope we have 1 $C_s$ and 228 $C_{ns}$ complementarity values, each of which is related to a specific patch pair. Normalizing over this set of numbers and looking at the Z-score regarding the specific one, we assessed the propensity that characterizes each antibody toward its specific antigen. In Figure \ref{fig:metodo_dataset}.C we report the distribution of such specific Z-scores. As evident, they are mostly negatives (86\% of the cases have a Z-score lower than 0, 28\% have a Z-score lower than -1), providing evidence that specific interactions are characterized by lower distances (higher complementarity) than non-specific ones.

Taken together, these results confirm the ability of our method to correctly capture the main determinant of molecular shape complementarity.

\subsection{Zernike based Monte Carlo simulation for molecular interface optimization}

\begin{figure*}[!]
\centering
\includegraphics[width =0.8 \textwidth]{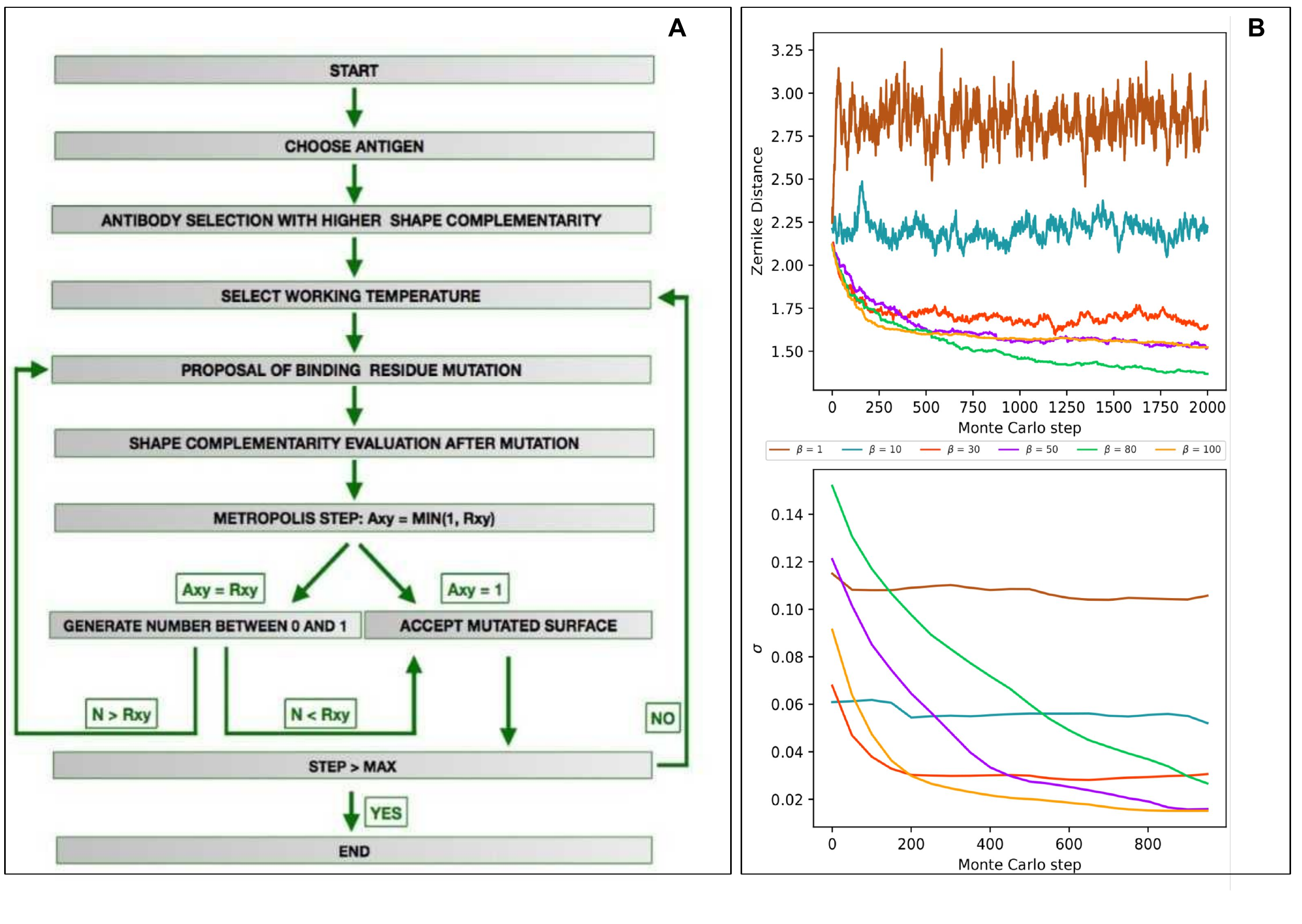}
\caption{\textbf{Development of a Monte Carlo for shape complementarity optimization against a target region}. \textbf{A)} Flowchart depicting the main steps of the computational protocol we developed. \textbf{B)}  Results of the mutagenesis Monte Carlo procedures performed at fixed $\beta$ ($\beta$ = 1, 10, 30, 50, 80, 100). The top panels represent the energy (i.e. the shape complementarity) as a function of the Monte Carlo steps. The low panels show the standard deviation of energy of the remaining part of the Monte Carlo simulations as a function of the Montecarlo steps (i.e. $\sigma(E)$ for n= 1000 means standard deviations of the energy obtained in the steps 1001-2000). 
}
\label{fig:flowchart}
\end{figure*}

The Zernike formalism enjoys several advantageous features in representing molecular surface: mainly, the invariance under rotation that makes such descriptors an "absolute" characterization of local protein morphology and the low computational cost of its calculation.  Indeed, in this section, we present our algorithm that, exploiting these advantages, aims to optimize the shape complementarity of an antibody toward a given molecular target region. A similar procedure has already been presented and tested in our previous work \cite{di2020novel}, and here for the first time, it is applied to antibody-antigen interaction systems.

Figure \ref{fig:flowchart}.A illustrates the main steps of the algorithm. We defined the target region as the portion of the antigen surface toward which an antibody will be optimized. It is thus necessary to identify the antibody chosen as a starting point of the algorithm. Summarized with the Zernike descriptors of the target region we compute the shape complementarity with all the paratopes of our dataset: here, the template, i.e. the antibody selected as the starting point for the mutagenesis process, can be chosen among the paratopes characterized by a high initial complementarity.

Established the template, we perform a Monte Carlo simulation employing computational mutagenesis on the paratope residues. In each step, we randomly select a residue mutating it in another of the 19 possible ones. The mutation generates a different paratope, characterized by a different shape of its molecular surface. Consequently, recomputing the Zernike descriptors we can evaluate the effect of the mutation on the complementarity with the target region: indeed we can define the \textit{complementarity balance} as:

\begin{equation}
    \Delta C =C_{mut} - C_{wt} =\\ D(p_{mut}, e_{tar}) - D(p_{orig}, e_{tar})
\end{equation}

where $p_{orig}$ and $p_{mut}$ are the Zernike descriptors of the original and the mutated paratope respectively, while the $e_{tar}$ represent the Zernike descriptors of the target epitope and $D$ represent the distance between 2 sets of descriptors. Since, as said, a high complementarity is reached when $D$ is low, $\Delta C < 0$ means a higher complementary surface, and $\Delta C > 0$ is obtained when the mutation is deleterious since it causes a worsening of the shape complementarity.

The number of combinations of possible mutations in an interacting region, composed usually of tens of residues, is huge. Therefore, to effectively sample the space of the possible mutants, we perform a Monte Carlo Metropolis simulation, iterating the procedure described above, where the mutation in each step is accepted according to the following rules:

\begin{equation}
    P = \begin{cases}
    1 \ \ \ \ \ \ \ \ \ if \ \ \  \ \Delta C< 0 \\
    e^{-\beta \Delta C} \ \ if \ \ \ \Delta C \ge 0
    \end{cases}
\label{eq:Prob_accept}
\end{equation}

where $\beta \sim \frac{1}{T}$ is the temperature factor that determines the probability of acceptance of a step worsening complementarity. Note that this aspect is crucial to properly explore a large number of different mutants: $\beta$ is thus progressively increased during the Monte Carlo simulation, progressively confining the system in an energy minimum in a simulated annealing process \cite{kirkpatrick1983optimization}. 

To observe how many steps are necessary to reach the equilibrium for each $\beta$, we preliminary ran several fixed-temperature Monte Carlo simulations.  We selected the epitope of an antigen structure in our dataset (PDB id: 1AR1) and, excluding its one, we choose as the starting template the most complementary paratope in the structural dataset. We thus performed six different Monte Carlo simulations, each for a different $\beta$, where the acceptation probability in each mutagenesis step is given by Eq. \ref{eq:Prob_accept}. Performing 10 independent simulations of M = 2000 steps for each $\beta$, the averaged results we obtained are summarized in Figure \ref{fig:flowchart}.B. In the top panel, we reported the energy (i.e. the complementarity) as a function of the Monte Carlo steps. In the low panel, we reported the standard deviation of the energy of the remaining part of the Monte Carlo simulations as a function of the number of steps (i.e. $\sigma(E)$ for m = 1000 means standard deviations of the energy obtained in the steps 1001-2000). As expected, for low values of $\beta$ (i.e. high temperature) the system lives in a condition of indifferent equilibrium, where whatever mutation has an equal likelihood to be accepted, independently from its effect on complementarity. When, on the contrary, $\beta$ is high (low temperature), the energy of the system rapidly decreases to a stationary local minimum. This trend is confirmed by looking at the stationary value of energies or, equivalently, noting that standard deviation tends progressively to zero. 
In the light of these results, in our protocol, we set N = 700 for each temperature. In this way, we preliminary allow the system to move away from the starting local conformation, thus freezing it in a new energy minimum, characterized by an increased shape complementarity with the target region.

\subsection{A Case of Study: Application to SARS-CoV-2 spike Protein}

\begin{figure*}[]
\centering
\includegraphics[width =1 \textwidth]{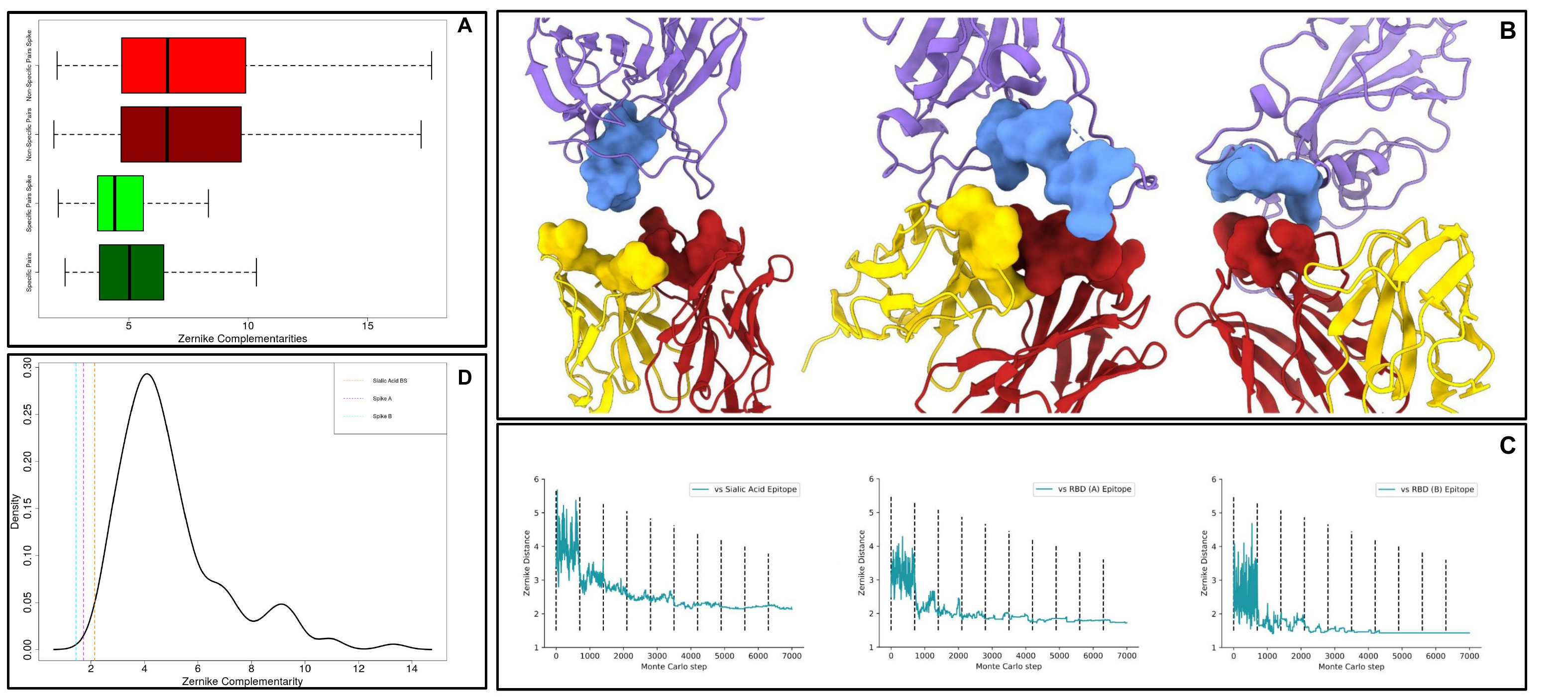}
\caption{ \textbf{Application of the optimization protocol to SARS-CoV-2 spike protein.} 
\textbf{A)} Boxplots comparing the \textit{specific complementarity} and the \textit{non-specific complementarity} in generic protein-antibody or in spike-antibody complexes. It is worth remarking that when the numerical value is low, the complementarity is high. \textbf{B)} Molecular representation of the optimized antibodies binding epitopes on spike protein. The antibody light and heavy chains are shown in yellow and red respectively, while the antigen is in purple. \textbf{C)} Shape complementarity as a function of the Monte Carlo steps for all the antibodies we optimized. Dashed lines separate different temperature intervals of the simulations.\textbf{D)} Probability density function of specific complementarities in the Spike dataset. The dashed lines represent the shape complementarity levels reached after the optimization protocols.}
\label{fig:applicazione_covid}
\end{figure*}

The approach described here is general and can be applied to whatever protein. This notwithstanding,  we applied it to the SARS-CoV-2 spike protein, a very relevant case of macromolecular interaction. Indeed, the severe acute respiratory syndrome coronavirus 2 infection is still causing very serious danger for public health \cite{huang2020clinical,zhu2020novel}.

Many therapeutic strategies are devoted to SARS-CoV-2 spike protein, protruding from the viral envelope and responsible for cell entry mechanism \cite{zhou2020pneumonia, walls2020structure,wan2020receptor}. Thus, we obtained, using the dedicated section of Cov-AbDab \cite{raybould2021cov}, a structural dataset of 145 spike-antibody complexes (we will call it the “Spike dataset”). We thus characterized the paratopes (antibody binding residues region) and the epitopes ( various regions on Spike) with the Zernike formalism. This allows us to compute, also for the Spike dataset, the \textit{specific complementarities} and the \textit{non-specific complementarities}, defined in Eq. \ref{eq:spec_dist} and \ref{eq:nonspec_dist}. The result of this analysis is shown in the following Figure \ref{fig:applicazione_covid}.A. These complementarities are reported as boxplots (light green for specific complementarities, light red for non-specific ones), even showing the boxplots regarding the general dataset (dark green and dark red, already shown in figure \ref{fig:metodo_dataset}.B). It results evident that, also in the Spike dataset, specific interactions are characterized by a complementarity much higher than non-specific interactions (k.s. test p-value < 2.2 e-16). As expected, the non-specific interactions are represented by very similar distributions, since in both the dataset they would represent an ensemble of non-interactions. Looking at these results, we noticed that the population of spike-Antibody complexes shows the same behavior as the general population of protein-antibody complexes.

In this framework, we selected on spike molecular surfaces three different regions as targets for the optimization protocol. On one hand, we targeted the two different molecular regions involved in the interaction between spike and angiotensin-converting enzyme 2  (ACE-2), the well-known cellular receptor responsible for viral cell invasion. Moreover, we optimized an antibody also toward a very exposed region in the N-terminal spike domain, responsible for contacting sialic acid molecules. Indeed, such interaction can confer to the virus, as occur for the Middle East respiratory syndrome coronavirus (MERS-CoV) \cite{li2017identification}, an additional molecular mechanism for cell intrusion.  The responsible spike region represents a promising therapeutic target \cite{milanetti2020silico, baker2020sars}.      

We selected the residues constituting such epitopes and we characterized their molecular surfaces through Zernike formalism. Thus, we calculated the complementarity between these regions and all the antibody binding sites in our original dataset. To begin the optimization from a favorite starting point, we selected as templates antibody binding sites characterized by the highest complementarity with each identified target.

We applied the procedure described in the previous section obtaining optimized paratopes whose molecular images are shown in Figure \ref{fig:applicazione_covid}.B, where we reported both the optimized antibodies and antigen interacting surfaces. In Figure \ref{fig:applicazione_covid}.C, we reported, for each of the Monte Carlo simulations performed, the shape complementarity as a function of the steps of the simulation, where the dashed lines enclose ranges with different $\beta$ values. Each simulation significantly optimizes shape complementarity, obtaining a Zernike distance decrease of 43\%, on average. Significantly, all the designed binding sites are characterized by a very high final shape complementarity, in terms of Zernike descriptors. Indeed, it is worth noting that the values obtained by all the three designed binding sites are lower than all the specific complementarities obtained in our structural dataset.

 Moreover, in Figure \ref{fig:applicazione_covid}.D, we reported the probability density function of specific interactions in the Spike dataset. The vertical dashed lines represent the final complementarity values we get after the optimization procedures. It has to be noted that our protocol can effectively optimize the shape complementarity, obtaining final shape complementarity similar to the best cases observed in the Spike dataset.

\begin{figure}[]
\includegraphics[width =0.8\columnwidth]{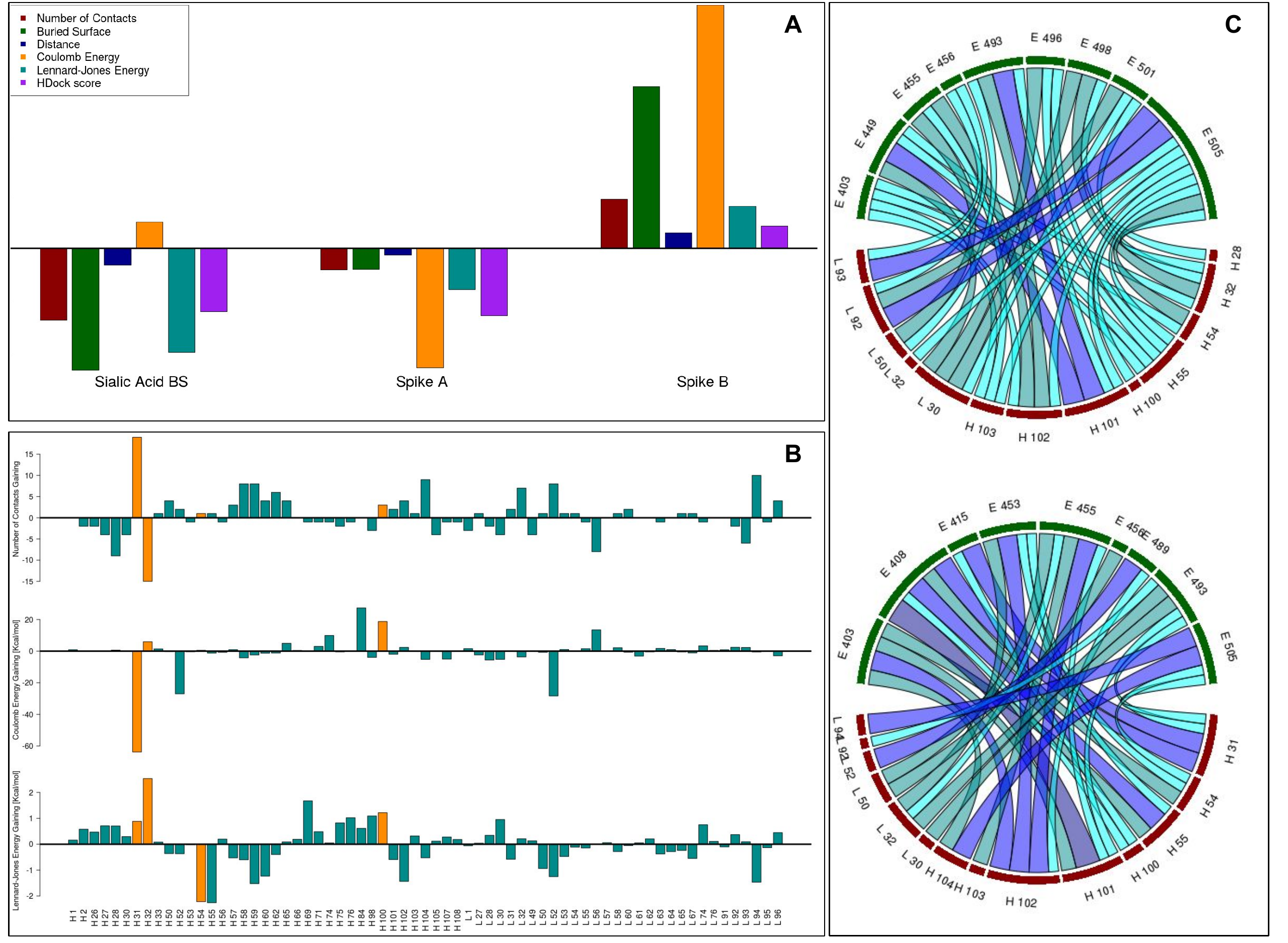}
\caption{ \textbf{Results of the docking analysis.}. \textbf{A)} Each bar represents the relative gaining (in terms of the number of residue-residue inter-molecular contacts, the surface buried in the complex, the mean inter-molecular distance of the closest atoms, the inter-molecular Coulomb energy, the inter-molecular Lennard-Jones energy, the HDock binding score) between the 10 best docking poses obtained with the original and the optimized antibodies. A positive value means an increase in binding compatibility. \textbf{B)} The gaining in terms of the number of intermolecular contacts, Coulomb energy, and Lennard-Jones energy each residue registered before and after the optimization. \textbf{C)} The network of residue-residue interactions at the interface when the original (upper figure) or the optimized (lower figure) antibody is docked to the spike B region. The color, from cyan to dark blue, and the width of the edges reflect the occurrences in the docking poses of a given contact. 
}
\label{fig:docking}
\end{figure}

The computational protocol we developed does not take into account several properties, known to be important in molecular recognition, such as electrostatics or hydrophobicity. 
In particular, our working hypothesis is that the shape complementarity plays a primary role as a perfect match between molecular surfaces is due to an optimal structural rearrangement, which is probably caused by the compatibility of amino acid compositions of the interacting patches. However, the relationship between shape complementarity and chemical-physical properties is not always trivial, requiring a further test for the patches proposed as interacting, to also analyze the compatibility of a chemical nature.
This means that a residue substitution can in principle worse the chemical compatibility between molecules, even if the shape complementarity is enhanced. For this reason, as a further step of our optimization protocol, we performed a molecular docking analysis using HDOCK \cite{yan2020hdock}. More specifically, we docked spike protein and the antibodies, both in the original and in the optimized versions, to study the effects our computational protocol has produced. We constrained docking to interact with the residues composing the spike target epitopes and the antibodies optimized regions. We summarized in Figure \ref{fig:docking} the results we obtained.

Thus, we selected the 10 best docking poses regarding both the original and the optimized antibodies. To assess whether the optimization protocol has been effective, some estimators of binding compatibility have been calculated. In particular, the number of residue-residue inter-molecular contacts, the surface buried in the complex, the mean distance of the closest atoms between the two interfaces, the Coulomb inter-molecular energy, the Lennard-Jones inter-molecular energy, the HDock binding score. For each of these observables, we computed the relative percentage of gaining after optimization so that positive values indicate an increased binding tightness (See Materials and Methods). As shown in Figure \ref{fig:docking}.A, even if in two applications we note a worsening, in one case the optimization procedure has produced an antibody with better values of all such estimators, indicating the importance of including the molecular docking approach as a filter of selected patches based on geometric compatibility.

We focused therefore on this case and we analyzed how the introduced residue substitutions were responsible for this better compatibility. Analyzing the best docking poses, in Figure \ref{fig:docking}.B we reported the gaining in terms of the number of intermolecular contacts and inter-molecular energy each residue registered before and after the optimization. Each residue is represented by a blue bar, while the residues mutated in the protocol are depicted in orange. As evident, the main effect regarding the residues "H 31" and "H 32". Indeed, to increase the shape complementarity, the optimization protocol preferred to switch the exposition of these residues. It can be noted that "H 31" residue, characterized by a high increase in the number of contacts, gains a very high amount of favorable (negative) Coulomb Energy. Moreover, even if the number of inter-molecular contacts gained by H 54 is negligible, such a residue ( and its neighborhood) acquire in the docking poses an increment of favorable Lennard-Jones energies. 

Lastly, we assessed the difference in residue-residue inter-molecular interaction networks. In Figure \ref{fig:docking}.C we reported the contacts between the main couples of residues, where a higher number of occurrences in the docking poses is testified by the thickness and the color of the edge. The spike residues are shown in green, the antibody ones are in red.
As further proof of the goodness of the proposed mutants, it can be noted that the interface of the optimized antibody (lower figure) is much more interconnected than the one of the original antibody (upper figure), indicating a possible effect on binding stability.

\section{Conclusions}

The binding affinity between biomolecules depends on a complex balance of several effects, including enthalpic and entropic contributions. Indeed, the substitution of even one residue at the interface could produce dramatic changes. Although many efforts were spent in this direction, predicting such effects has proven to be a difficult task and is still an open problem in computational biology. 

In this scenario, the evaluation of shape complementarity between molecular regions is undoubtedly a central aspect. In this work, we focused on antibody-antigen interaction, a relevant case of molecular recognition. We applied our recently developed formalism based on the 2D Zernike polynomials to evaluate the shape complementarity with a quantitative approach. Once summarized the topological properties of interacting regions with a set of numerical descriptors, we demonstrated that such formalism assigns to pairs of interacting regions complementarities statistically higher than the ones assigned to regions not in interaction.

We thus developed a Monte Carlo-based approach for the shape optimization of an antibody towards a molecular target region. We propose a new strategy that, potentially, can modify an antibody in order to acquire a very high shape complementarity for a given epitope of any antigen protein. 

Because of the emergence of viral variants that can eventually escape antibodies maturated in vaccinated or recovered patients, the interactions between antibodies and SARS-CoV-2 spike protein are being extensively studied and still needs further investigation.

For this reason, we selected three molecular regions on spike protein as the target epitopes for our procedure. We, therefore, devised a set of antibodies characterized by a high shape complementarity toward their cognate epitopes.

However, even without considering therapeutically important elements such as immunogenicity and solubility, some other aspects have to be properly considered in our algorithm to increase the probability of identifying actually binding antibodies. Firstly, to produce more reliable mutants structures, the residue substitutions procedure has to account for hypervariable loops canonical structure modeling. Moreover, we worked on antibodies bound structures: a structural conformational exploration can allow the antibodies to energetically rearrange their side chains, to find the proper conformation able to bind the studied antigen. Finally, the binding compatibility does not depend only on shape complementarity, thus the inclusion of terms accounting for residues' chemical characteristics surely will improve the method's performance.  

In conclusion, this procedure can represent a promising strategy for interface region molecular optimization, where the inclusion of the aspects discussed above represents the necessary improvement steps. In the present work, we highlighted with an independent molecular docking evaluation the case when the optimization procedure has increased molecular complex compactness.  
  

\section{Materials and Methods}

\subsection{Dataset}
We selected 229 protein-binding antibodies with sequence identity lower than 90\% and resolution $< 3.0 $ Å using the SabDab database \cite{dunbar2013sabdab}. 
The Spike dataset, i.e. the structural dataset of spike-antibody complexes was built using CoV-AbDab \cite{raybould2021cov}. It results in 145 complex structures with a sequence identity lower than 90\%, as calculated using CD-HIT\cite{huang2010cd}.

The sequence of each antibody was renumbered according to the Chothia numbering scheme \cite{chothia1987canonical,chothia1989conformations} using an in-house python script.

The structure of the SARS-CoV-2 spike protein used for the identification of the ACE2 interacting region has the PDB code 6vw1. When we investigated the N-terminal domain we used the structure 7jji.

We identified on spike protein two epitopes in the ACE2 binding region: spike A and spike B. spike A epitope is constituted by the residues: "TYR 453, LEU 455, PHE 456, ALA 475, GLY 476, PHE 486, ASN 487, TYR 489, GLN 493". spike B epitope is constituted by the residues: "TYR 449, GLY 496, GLN 498, THR 500, ASN 501, GLY 502, TYR 505". We identified another epitope in the N-terminal domain, in the region involved in sialic acid-binding. That region is defined as the set of residues whose CA atoms are closer than $8 \AA$ to the TRP 258 CA \cite{milanetti2020silico}. Such epitope is constituted by these residues: "LEU 244, HIS 245, ARG 246, SER 247, TYR 248, LEU 249, THR 250, PRO 251, GLY 252, ASP 253, SER 254, SER 255, SER 256, GLY 257, TRP 258, THR 259, ALA 260".

\subsection{Surface Construction}

Using as reference the experimental structures, computational mutagenesis has been performed using SCWRL4 \cite{krivov2009improved}.

For each protein structure, Solvent Accessible Surface is computed using DMS software with standard option \cite{richards1977areas}. The interacting surface is constituted by the surface points belonging to interacting residues, defined as the set of residues having at least one atom closer than 4 $\AA$ to any atoms of the molecular partner. 

\subsection{Zernike Descriptors}

Given a 2D function $f(r,\phi)$ in the unitary circle( region $r < 1$), it can be expanded in the Zernike polynomials basis. Therefore:

\begin{equation}
f(r,\phi) = \sum_{n=0}^{\infty} \sum_{m=0}^{m=n} c_{nm} Z_{nm}
\end{equation}

where  

\begin{equation}
\begin{split}
        c_{nm} =&  \frac{(n+1)}{\pi} \left<{Z_{nm}|f}\right>= \\
        &= \frac{(n+1)}{\pi} \int_0^1 dr r\int_0^{2\pi} d\phi Z_{nm}^*(r,\phi) f(r,\phi)
    \end{split}
\end{equation}

are the expansion coefficients (Zernike moments). The complex functions $Z_{nm}(r, \phi)$ are the Zernike polynomials, each composed of a radial and an angular part:

\begin{equation}
Z_{nm} = R_{nm}(r) e^{im\phi}.
\end{equation}

The radial dependence, given $n$ and $m$, can be written as follow:

\begin{equation}
R_{nm}(r) = 
\sum_{k= 0}^{\frac{n-m}{2}} \frac{(-1)^k (n-k)!}{k!\left(\frac{n+m}{2} - k\right)!\left(\frac{n-m}{2}-k\right)!} r^{n-2k}
\end{equation}

For each couple of polynomials, this rule holds:

\begin{equation}
\left<{Z_{nm}|Z_{n'm'}}\right> = \frac{\pi }{(n+1)}\delta_{nn'}\delta_{mm'}
\end{equation}

Therefore the set of polynomials forms a basis. Knowing the coefficients, $\{ c_{nm} \}$ allows the reconstruction of the original function. The level of the detail can be modified by modulating the order of expansion, $N = \text{max(n)}$.

The norm of each coefficient ($z_{nm} = |c_{nm}|$) does not depend on the phase, therefore it is invariant under rotations around the origin. 

The shape complementarity between two regions can be evaluated by comparing their Zernike invariants. In particular, we  measured the complementarity between region $i$ and $j$ as the Euclidean distance between the invariant vectors, i.e.
\begin{equation} 
\label{eq:d}
d_{ij} = \sqrt{\sum_{k=1}^{M=121} (z_i^k - z_j^k)^2}
\end{equation}

We adopted N=20, therefore dealing with 121 invariant descriptors for each patch.

\subsection{Analysis of docking poses}

\begin{table}[!ht]
    \centering
    \begin{tabular}{||l|l|l|l|l|l||}
    \hline
        Epitope & Chain ID & Resno & Insert & Original Res & Inserted Res \\ \hline
        Sialic Acid BS & H & 33 & ~ & TYR & TRP \\ \hline
        Sialic Acid BS ~ & H & 52 & ~ & TYR & TRP \\ \hline
        Sialic Acid BS~ & H & 95 & ~ & TRP & GLN \\ \hline
        Sialic Acid BS ~ & H & 100 & S & PHE & ARG \\ \hline
        Sialic Acid BS ~ & L & 30 & B & SER & ALA \\ \hline
        Sialic Acid BS ~ & L & 30 & C & ILE & PHE \\ \hline
        Sialic Acid BS ~ & L & 30 & D & SER & ILE \\ \hline
        Sialic Acid BS ~ & L & 31 & ~ & TYR & ALA \\ \hline
        Sialic Acid BS ~ & L & 50 & ~ & TRP & GLY \\ \hline
        Sialic Acid BS ~ & L & 91 & ~ & HIS & SER \\ \hline
        Sialic Acid BS ~ & L & 92 & ~ & TYR & GLY \\ \hline
        Sialic Acid BS ~ & L & 93 & ~ & THR & HIS \\ \hline
        Sialic Acid BS ~ & L & 94 & ~ & THR & VAL \\ \hline
        \hline
        spike A & H & 31 & ~ & SER & LYS \\ \hline
        spike A ~ & H & 96 & ~ & HIS & MET \\ \hline
        spike A ~ & H & 97 & ~ & TYR & PHE \\ \hline
        spike A ~ & H & 98 & ~ & GLY & LEU \\ \hline
        spike A ~ & H & 99 & ~ & LEU & PRO \\ \hline
        spike A ~ & H & 100 & S & ASP & HIS \\ \hline
        spike A ~ & H & 100 & T & TRP & LYS \\ \hline
        spike A ~ & L & 30 & ~ & TYR & ILE \\ \hline
        spike A ~ & L & 32 & ~ & TRP & VAL \\ \hline
        spike A ~ & L & 91 & ~ & GLY & ARG \\ \hline
        spike A ~ & L & 92 & ~ & GLN & ARG \\ \hline
        \hline
        spike B & H & 31 & ~ & GLY & GLU \\ \hline
        spike B & H & 32 & ~ & TYR & CYS \\ \hline
        spike B & H & 53 & ~ & TYR & PHE \\ \hline
        \end{tabular}
    \caption{The residue substitutions performed during shape optimization procedure. The template structures for Sialic Acid Binding Site, spike A and spike B were 3bdy, 1yjd, 1kb5, respectively. We adopted the Chotia numbering scheme.}
    \label{tab:results}
\end{table}

In the table \ref{tab:results}, we report the mutations proposed as a result of the three Monte Carlo simulations.

We docked the three original and the three optimized antibody structures with spike using HDOCK\cite{yan2020hdock}, indicating as interacting residues the one written in the Dataset section. 

We selected, for all the 6 docking simulations, the best 10 poses according to the Hdock binding score, an iterative knowledge-based scoring function. For each pose we get:
\begin{itemize}
    \item the number of inter-molecular residue-residue contacts. Two residues are in contact if the minimum distance between their atoms is less than 4 $\AA$.
    \item the surface buried in the complex. The surface buried is defined as the difference between the sum of the monomers' area and the complex area. For this calculation we use DMS software \cite{richards1977areas}.
    \item the mean of the lowest 100 atom-atom inter-molecular distances. 
    \item the sum of the Coulomb energy of the interactions occurring between antibody and spike atoms. We used CHARMM27 force field \cite{mackerell2002charmm}. 
    \item the sum of the Lennard-Jones energy of the interactions occurring between antibody and spike atoms. We used CHARMM27 force field \cite{mackerell2002charmm}. 
    \item the pose Hdock binding score
\end{itemize}

The comparisons between the results regarding original and optimized antibodies are performed so as a positive value means an increase in binding compatibility after optimization. Therefore the relative percentage of gaining is defined as:
\begin{itemize}
    \item Number of contacts :$ \frac{<Cont>_{opt} - <Cont>_{orig}}{<Cont>_{orig}} $
    \item Buried Area :$ \frac{<Surf>_{opt} - <Surf>_{orig}}{<Surf>_{orig}} $
    \item Distance :$ \frac{<Dist>_{orig} - <Dist>_{opt}}{<dist>_{orig}} $
    \item Coulomb Energy :$ \frac{<E_c>_{orig} - <E_c>_{opt}}{<E_c>_{orig}} $
    \item Lennard-Jones Energy :$ \frac{<E_{lj}>_{orig} - <E_lj>_{opt}}{<E_lj>_{orig}} $
    \item HDock score :$ \frac{<Score>_{orig} - <Score_c>_{opt}}{<Score>_{orig}} $
\end{itemize}

where the subscripts "orig" and "opt" refer to the poses obtained with antibodies before and after the optimization, respectively.

\section*{Data Availability Statement}

All the molecular structures used for this work are available on Protein Data Bank (https://www.rcsb.org/). 

\section*{Conflict of Interest Disclosure}

The authors declare no conflicts of interest.

\section*{Acknowledgments}
The research leading to these results has been also supported by European Research Council Synergy grant ASTRA (n. 855923).


\begin{thebibliography}{}
\expandafter\ifx\csname natexlab\endcsname\relax\def\natexlab#1{#1}\fi
\expandafter\ifx\csname bibnamefont\endcsname\relax
  \def\bibnamefont#1{#1}\fi
\expandafter\ifx\csname bibfnamefont\endcsname\relax
  \def\bibfnamefont#1{#1}\fi
\expandafter\ifx\csname citenamefont\endcsname\relax
  \def\citenamefont#1{#1}\fi
\expandafter\ifx\csname url\endcsname\relax
  \def\url#1{\texttt{#1}}\fi
\expandafter\ifx\csname urlprefix\endcsname\relax\def\urlprefix{URL }\fi
\providecommand{\bibinfo}[2]{#2}
\providecommand{\eprint}[2][]{\url{#2}}

\bibitem[{\citenamefont{Jones and Thornton}(1996)}]{jones1996principles}
\bibinfo{author}{\bibfnamefont{S.}~\bibnamefont{Jones}} \bibnamefont{and}
  \bibinfo{author}{\bibfnamefont{J.~M.} \bibnamefont{Thornton}},
  \bibinfo{journal}{Proceedings of the National Academy of Sciences}
  \textbf{\bibinfo{volume}{93}}, \bibinfo{pages}{13} (\bibinfo{year}{1996}).

\bibitem[{\citenamefont{Gromiha et~al.}(2017)\citenamefont{Gromiha, Yugandhar,
  and Jemimah}}]{gromiha2017protein}
\bibinfo{author}{\bibfnamefont{M.~M.} \bibnamefont{Gromiha}},
  \bibinfo{author}{\bibfnamefont{K.}~\bibnamefont{Yugandhar}},
  \bibnamefont{and} \bibinfo{author}{\bibfnamefont{S.}~\bibnamefont{Jemimah}},
  \bibinfo{journal}{Current opinion in structural biology}
  \textbf{\bibinfo{volume}{44}}, \bibinfo{pages}{31} (\bibinfo{year}{2017}).

\bibitem[{\citenamefont{Gavin et~al.}(2002)\citenamefont{Gavin, B{\"o}sche,
  Krause, Grandi, Marzioch, Bauer, Schultz, Rick, Michon, Cruciat
  et~al.}}]{gavin2002functional}
\bibinfo{author}{\bibfnamefont{A.-C.} \bibnamefont{Gavin}},
  \bibinfo{author}{\bibfnamefont{M.}~\bibnamefont{B{\"o}sche}},
  \bibinfo{author}{\bibfnamefont{R.}~\bibnamefont{Krause}},
  \bibinfo{author}{\bibfnamefont{P.}~\bibnamefont{Grandi}},
  \bibinfo{author}{\bibfnamefont{M.}~\bibnamefont{Marzioch}},
  \bibinfo{author}{\bibfnamefont{A.}~\bibnamefont{Bauer}},
  \bibinfo{author}{\bibfnamefont{J.}~\bibnamefont{Schultz}},
  \bibinfo{author}{\bibfnamefont{J.~M.} \bibnamefont{Rick}},
  \bibinfo{author}{\bibfnamefont{A.-M.} \bibnamefont{Michon}},
  \bibinfo{author}{\bibfnamefont{C.-M.} \bibnamefont{Cruciat}},
  \bibnamefont{et~al.}, \bibinfo{journal}{Nature}
  \textbf{\bibinfo{volume}{415}}, \bibinfo{pages}{141} (\bibinfo{year}{2002}).

\bibitem[{\citenamefont{Han et~al.}(2004)\citenamefont{Han, Bertin, Hao,
  Goldberg, Berriz, Zhang, Dupuy, Walhout, Cusick, Roth
  et~al.}}]{han2004evidence}
\bibinfo{author}{\bibfnamefont{J.-D.~J.} \bibnamefont{Han}},
  \bibinfo{author}{\bibfnamefont{N.}~\bibnamefont{Bertin}},
  \bibinfo{author}{\bibfnamefont{T.}~\bibnamefont{Hao}},
  \bibinfo{author}{\bibfnamefont{D.~S.} \bibnamefont{Goldberg}},
  \bibinfo{author}{\bibfnamefont{G.~F.} \bibnamefont{Berriz}},
  \bibinfo{author}{\bibfnamefont{L.~V.} \bibnamefont{Zhang}},
  \bibinfo{author}{\bibfnamefont{D.}~\bibnamefont{Dupuy}},
  \bibinfo{author}{\bibfnamefont{A.~J.} \bibnamefont{Walhout}},
  \bibinfo{author}{\bibfnamefont{M.~E.} \bibnamefont{Cusick}},
  \bibinfo{author}{\bibfnamefont{F.~P.} \bibnamefont{Roth}},
  \bibnamefont{et~al.}, \bibinfo{journal}{Nature}
  \textbf{\bibinfo{volume}{430}}, \bibinfo{pages}{88} (\bibinfo{year}{2004}).

\bibitem[{\citenamefont{Gainza et~al.}(2020)\citenamefont{Gainza, Sverrisson,
  Monti, Rodola, Boscaini, Bronstein, and Correia}}]{gainza2020deciphering}
\bibinfo{author}{\bibfnamefont{P.}~\bibnamefont{Gainza}},
  \bibinfo{author}{\bibfnamefont{F.}~\bibnamefont{Sverrisson}},
  \bibinfo{author}{\bibfnamefont{F.}~\bibnamefont{Monti}},
  \bibinfo{author}{\bibfnamefont{E.}~\bibnamefont{Rodola}},
  \bibinfo{author}{\bibfnamefont{D.}~\bibnamefont{Boscaini}},
  \bibinfo{author}{\bibfnamefont{M.}~\bibnamefont{Bronstein}},
  \bibnamefont{and} \bibinfo{author}{\bibfnamefont{B.}~\bibnamefont{Correia}},
  \bibinfo{journal}{Nature Methods} \textbf{\bibinfo{volume}{17}},
  \bibinfo{pages}{184} (\bibinfo{year}{2020}).

\bibitem[{\citenamefont{Milanetti
  et~al.}(2021{\natexlab{a}})\citenamefont{Milanetti, Miotto, Di~Rienzo, Monti,
  Gosti, and Ruocco}}]{milanetti20212d}
\bibinfo{author}{\bibfnamefont{E.}~\bibnamefont{Milanetti}},
  \bibinfo{author}{\bibfnamefont{M.}~\bibnamefont{Miotto}},
  \bibinfo{author}{\bibfnamefont{L.}~\bibnamefont{Di~Rienzo}},
  \bibinfo{author}{\bibfnamefont{M.}~\bibnamefont{Monti}},
  \bibinfo{author}{\bibfnamefont{G.}~\bibnamefont{Gosti}}, \bibnamefont{and}
  \bibinfo{author}{\bibfnamefont{G.}~\bibnamefont{Ruocco}},
  \bibinfo{journal}{Computational and structural biotechnology journal}
  \textbf{\bibinfo{volume}{19}}, \bibinfo{pages}{29}
  (\bibinfo{year}{2021}{\natexlab{a}}).

\bibitem[{\citenamefont{Siebenmorgen and
  Zacharias}(2020)}]{siebenmorgen2020computational}
\bibinfo{author}{\bibfnamefont{T.}~\bibnamefont{Siebenmorgen}}
  \bibnamefont{and}
  \bibinfo{author}{\bibfnamefont{M.}~\bibnamefont{Zacharias}},
  \bibinfo{journal}{Wiley Interdisciplinary Reviews: Computational Molecular
  Science} \textbf{\bibinfo{volume}{10}}, \bibinfo{pages}{e1448}
  (\bibinfo{year}{2020}).

\bibitem[{\citenamefont{Vangone and Bonvin}(2015)}]{vangone2015contacts}
\bibinfo{author}{\bibfnamefont{A.}~\bibnamefont{Vangone}} \bibnamefont{and}
  \bibinfo{author}{\bibfnamefont{A.~M.} \bibnamefont{Bonvin}},
  \bibinfo{journal}{elife} \textbf{\bibinfo{volume}{4}},
  \bibinfo{pages}{e07454} (\bibinfo{year}{2015}).

\bibitem[{\citenamefont{Vakser}(2014)}]{vakser2014protein}
\bibinfo{author}{\bibfnamefont{I.~A.} \bibnamefont{Vakser}},
  \bibinfo{journal}{Biophysical journal} \textbf{\bibinfo{volume}{107}},
  \bibinfo{pages}{1785} (\bibinfo{year}{2014}).

\bibitem[{\citenamefont{Kozakov et~al.}(2017)\citenamefont{Kozakov, Hall, Xia,
  Porter, Padhorny, Yueh, Beglov, and Vajda}}]{kozakov2017cluspro}
\bibinfo{author}{\bibfnamefont{D.}~\bibnamefont{Kozakov}},
  \bibinfo{author}{\bibfnamefont{D.~R.} \bibnamefont{Hall}},
  \bibinfo{author}{\bibfnamefont{B.}~\bibnamefont{Xia}},
  \bibinfo{author}{\bibfnamefont{K.~A.} \bibnamefont{Porter}},
  \bibinfo{author}{\bibfnamefont{D.}~\bibnamefont{Padhorny}},
  \bibinfo{author}{\bibfnamefont{C.}~\bibnamefont{Yueh}},
  \bibinfo{author}{\bibfnamefont{D.}~\bibnamefont{Beglov}}, \bibnamefont{and}
  \bibinfo{author}{\bibfnamefont{S.}~\bibnamefont{Vajda}},
  \bibinfo{journal}{Nature protocols} \textbf{\bibinfo{volume}{12}},
  \bibinfo{pages}{255} (\bibinfo{year}{2017}).

\bibitem[{\citenamefont{Geng et~al.}(2020)\citenamefont{Geng, Jung, Renaud,
  Honavar, Bonvin, and Xue}}]{geng2020iscore}
\bibinfo{author}{\bibfnamefont{C.}~\bibnamefont{Geng}},
  \bibinfo{author}{\bibfnamefont{Y.}~\bibnamefont{Jung}},
  \bibinfo{author}{\bibfnamefont{N.}~\bibnamefont{Renaud}},
  \bibinfo{author}{\bibfnamefont{V.}~\bibnamefont{Honavar}},
  \bibinfo{author}{\bibfnamefont{A.~M.} \bibnamefont{Bonvin}},
  \bibnamefont{and} \bibinfo{author}{\bibfnamefont{L.~C.} \bibnamefont{Xue}},
  \bibinfo{journal}{Bioinformatics} \textbf{\bibinfo{volume}{36}},
  \bibinfo{pages}{112} (\bibinfo{year}{2020}).

\bibitem[{\citenamefont{Yan et~al.}(2020)\citenamefont{Yan, Tao, He, and
  Huang}}]{yan2020hdock}
\bibinfo{author}{\bibfnamefont{Y.}~\bibnamefont{Yan}},
  \bibinfo{author}{\bibfnamefont{H.}~\bibnamefont{Tao}},
  \bibinfo{author}{\bibfnamefont{J.}~\bibnamefont{He}}, \bibnamefont{and}
  \bibinfo{author}{\bibfnamefont{S.-Y.} \bibnamefont{Huang}},
  \bibinfo{journal}{Nature protocols} \textbf{\bibinfo{volume}{15}},
  \bibinfo{pages}{1829} (\bibinfo{year}{2020}).

\bibitem[{\citenamefont{Katchalski-Katzir
  et~al.}(1992)\citenamefont{Katchalski-Katzir, Shariv, Eisenstein, Friesem,
  Aflalo, and Vakser}}]{katchalski1992molecular}
\bibinfo{author}{\bibfnamefont{E.}~\bibnamefont{Katchalski-Katzir}},
  \bibinfo{author}{\bibfnamefont{I.}~\bibnamefont{Shariv}},
  \bibinfo{author}{\bibfnamefont{M.}~\bibnamefont{Eisenstein}},
  \bibinfo{author}{\bibfnamefont{A.~A.} \bibnamefont{Friesem}},
  \bibinfo{author}{\bibfnamefont{C.}~\bibnamefont{Aflalo}}, \bibnamefont{and}
  \bibinfo{author}{\bibfnamefont{I.~A.} \bibnamefont{Vakser}},
  \bibinfo{journal}{Proceedings of the National Academy of Sciences}
  \textbf{\bibinfo{volume}{89}}, \bibinfo{pages}{2195} (\bibinfo{year}{1992}).

\bibitem[{\citenamefont{Lawrence and Colman}(1993)}]{lawrence1993shape}
\bibinfo{author}{\bibfnamefont{M.~C.} \bibnamefont{Lawrence}} \bibnamefont{and}
  \bibinfo{author}{\bibfnamefont{P.~M.} \bibnamefont{Colman}},
  \emph{\bibinfo{title}{Shape complementarity at protein/protein interfaces}}
  (\bibinfo{year}{1993}).

\bibitem[{\citenamefont{Chen and Weng}(2003)}]{chen2003novel}
\bibinfo{author}{\bibfnamefont{R.}~\bibnamefont{Chen}} \bibnamefont{and}
  \bibinfo{author}{\bibfnamefont{Z.}~\bibnamefont{Weng}},
  \bibinfo{journal}{Proteins: Structure, Function, and Bioinformatics}
  \textbf{\bibinfo{volume}{51}}, \bibinfo{pages}{397} (\bibinfo{year}{2003}).

\bibitem[{\citenamefont{Nicola and Vakser}(2007)}]{nicola2007simple}
\bibinfo{author}{\bibfnamefont{G.}~\bibnamefont{Nicola}} \bibnamefont{and}
  \bibinfo{author}{\bibfnamefont{I.~A.} \bibnamefont{Vakser}},
  \bibinfo{journal}{Bioinformatics} \textbf{\bibinfo{volume}{23}},
  \bibinfo{pages}{789} (\bibinfo{year}{2007}).

\bibitem[{\citenamefont{Kuroda and Gray}(2016)}]{kuroda2016shape}
\bibinfo{author}{\bibfnamefont{D.}~\bibnamefont{Kuroda}} \bibnamefont{and}
  \bibinfo{author}{\bibfnamefont{J.~J.} \bibnamefont{Gray}},
  \bibinfo{journal}{Bioinformatics} \textbf{\bibinfo{volume}{32}},
  \bibinfo{pages}{2451} (\bibinfo{year}{2016}).

\bibitem[{\citenamefont{Yan and Huang}(2019)}]{yan2019pushing}
\bibinfo{author}{\bibfnamefont{Y.}~\bibnamefont{Yan}} \bibnamefont{and}
  \bibinfo{author}{\bibfnamefont{S.-Y.} \bibnamefont{Huang}},
  \bibinfo{journal}{BMC bioinformatics} \textbf{\bibinfo{volume}{20}},
  \bibinfo{pages}{1} (\bibinfo{year}{2019}).

\bibitem[{\citenamefont{Erijman et~al.}(2014)\citenamefont{Erijman, Rosenthal,
  and Shifman}}]{erijman2014structure}
\bibinfo{author}{\bibfnamefont{A.}~\bibnamefont{Erijman}},
  \bibinfo{author}{\bibfnamefont{E.}~\bibnamefont{Rosenthal}},
  \bibnamefont{and} \bibinfo{author}{\bibfnamefont{J.~M.}
  \bibnamefont{Shifman}}, \bibinfo{journal}{PLOS one}
  \textbf{\bibinfo{volume}{9}}, \bibinfo{pages}{e110085}
  (\bibinfo{year}{2014}).

\bibitem[{\citenamefont{Venkatraman
  et~al.}(2009{\natexlab{a}})\citenamefont{Venkatraman, Sael, and
  Kihara}}]{venkatraman2009potential}
\bibinfo{author}{\bibfnamefont{V.}~\bibnamefont{Venkatraman}},
  \bibinfo{author}{\bibfnamefont{L.}~\bibnamefont{Sael}}, \bibnamefont{and}
  \bibinfo{author}{\bibfnamefont{D.}~\bibnamefont{Kihara}},
  \bibinfo{journal}{Cell biochemistry and biophysics}
  \textbf{\bibinfo{volume}{54}}, \bibinfo{pages}{23}
  (\bibinfo{year}{2009}{\natexlab{a}}).

\bibitem[{\citenamefont{Di~Rienzo et~al.}(2017)\citenamefont{Di~Rienzo,
  Milanetti, Lepore, Olimpieri, and Tramontano}}]{di2017superposition}
\bibinfo{author}{\bibfnamefont{L.}~\bibnamefont{Di~Rienzo}},
  \bibinfo{author}{\bibfnamefont{E.}~\bibnamefont{Milanetti}},
  \bibinfo{author}{\bibfnamefont{R.}~\bibnamefont{Lepore}},
  \bibinfo{author}{\bibfnamefont{P.~P.} \bibnamefont{Olimpieri}},
  \bibnamefont{and}
  \bibinfo{author}{\bibfnamefont{A.}~\bibnamefont{Tramontano}},
  \bibinfo{journal}{Scientific reports} \textbf{\bibinfo{volume}{7}},
  \bibinfo{pages}{1} (\bibinfo{year}{2017}).

\bibitem[{\citenamefont{Di~Rienzo
  et~al.}(2021{\natexlab{a}})\citenamefont{Di~Rienzo, Milanetti, Ruocco, and
  Lepore}}]{di2021quantitative}
\bibinfo{author}{\bibfnamefont{L.}~\bibnamefont{Di~Rienzo}},
  \bibinfo{author}{\bibfnamefont{E.}~\bibnamefont{Milanetti}},
  \bibinfo{author}{\bibfnamefont{G.}~\bibnamefont{Ruocco}}, \bibnamefont{and}
  \bibinfo{author}{\bibfnamefont{R.}~\bibnamefont{Lepore}},
  \bibinfo{journal}{Frontiers in molecular biosciences} p. \bibinfo{pages}{933}
  (\bibinfo{year}{2021}{\natexlab{a}}).

\bibitem[{\citenamefont{Daberdaku and Ferrari}(2019)}]{daberdaku2019antibody}
\bibinfo{author}{\bibfnamefont{S.}~\bibnamefont{Daberdaku}} \bibnamefont{and}
  \bibinfo{author}{\bibfnamefont{C.}~\bibnamefont{Ferrari}},
  \bibinfo{journal}{Bioinformatics} \textbf{\bibinfo{volume}{35}},
  \bibinfo{pages}{1870} (\bibinfo{year}{2019}).

\bibitem[{\citenamefont{Zernike}(1934)}]{zernike1934diffraction}
\bibinfo{author}{\bibfnamefont{F.}~\bibnamefont{Zernike}},
  \bibinfo{journal}{Monthly Notices of the Royal Astronomical Society}
  \textbf{\bibinfo{volume}{94}}, \bibinfo{pages}{377} (\bibinfo{year}{1934}).

\bibitem[{\citenamefont{Novotni and Klein}(2004)}]{novotni2004shape}
\bibinfo{author}{\bibfnamefont{M.}~\bibnamefont{Novotni}} \bibnamefont{and}
  \bibinfo{author}{\bibfnamefont{R.}~\bibnamefont{Klein}},
  \bibinfo{journal}{Computer-Aided Design} \textbf{\bibinfo{volume}{36}},
  \bibinfo{pages}{1047} (\bibinfo{year}{2004}).

\bibitem[{\citenamefont{Canterakis}(1999)}]{canterakis19993d}
\bibinfo{author}{\bibfnamefont{N.}~\bibnamefont{Canterakis}}, in
  \emph{\bibinfo{booktitle}{In 11th Scandinavian Conf. on Image Analysis}}
  (\bibinfo{organization}{Citeseer}, \bibinfo{year}{1999}).

\bibitem[{\citenamefont{Daberdaku and Ferrari}(2018)}]{daberdaku2018exploring}
\bibinfo{author}{\bibfnamefont{S.}~\bibnamefont{Daberdaku}} \bibnamefont{and}
  \bibinfo{author}{\bibfnamefont{C.}~\bibnamefont{Ferrari}},
  \bibinfo{journal}{BMC bioinformatics} \textbf{\bibinfo{volume}{19}},
  \bibinfo{pages}{35} (\bibinfo{year}{2018}).

\bibitem[{\citenamefont{Di~Rienzo
  et~al.}(2020{\natexlab{a}})\citenamefont{Di~Rienzo, Milanetti, Alba, and
  D’Abramo}}]{di2020quantitative}
\bibinfo{author}{\bibfnamefont{L.}~\bibnamefont{Di~Rienzo}},
  \bibinfo{author}{\bibfnamefont{E.}~\bibnamefont{Milanetti}},
  \bibinfo{author}{\bibfnamefont{J.}~\bibnamefont{Alba}}, \bibnamefont{and}
  \bibinfo{author}{\bibfnamefont{M.}~\bibnamefont{D’Abramo}},
  \bibinfo{journal}{Journal of Chemical Information and Modeling}
  \textbf{\bibinfo{volume}{60}}, \bibinfo{pages}{1390}
  (\bibinfo{year}{2020}{\natexlab{a}}).

\bibitem[{\citenamefont{Venkatraman
  et~al.}(2009{\natexlab{b}})\citenamefont{Venkatraman, Yang, Sael, and
  Kihara}}]{venkatraman2009protein}
\bibinfo{author}{\bibfnamefont{V.}~\bibnamefont{Venkatraman}},
  \bibinfo{author}{\bibfnamefont{Y.~D.} \bibnamefont{Yang}},
  \bibinfo{author}{\bibfnamefont{L.}~\bibnamefont{Sael}}, \bibnamefont{and}
  \bibinfo{author}{\bibfnamefont{D.}~\bibnamefont{Kihara}},
  \bibinfo{journal}{BMC bioinformatics} \textbf{\bibinfo{volume}{10}},
  \bibinfo{pages}{407} (\bibinfo{year}{2009}{\natexlab{b}}).

\bibitem[{\citenamefont{Kihara et~al.}(2011)\citenamefont{Kihara, Sael, Chikhi,
  and Esquivel-Rodriguez}}]{kihara2011molecular}
\bibinfo{author}{\bibfnamefont{D.}~\bibnamefont{Kihara}},
  \bibinfo{author}{\bibfnamefont{L.}~\bibnamefont{Sael}},
  \bibinfo{author}{\bibfnamefont{R.}~\bibnamefont{Chikhi}}, \bibnamefont{and}
  \bibinfo{author}{\bibfnamefont{J.}~\bibnamefont{Esquivel-Rodriguez}},
  \bibinfo{journal}{Current Protein and Peptide Science}
  \textbf{\bibinfo{volume}{12}}, \bibinfo{pages}{520} (\bibinfo{year}{2011}).

\bibitem[{\citenamefont{Alba et~al.}(2020)\citenamefont{Alba, Di~Rienzo,
  Milanetti, Acuto, and D’Abramo}}]{alba2020molecular}
\bibinfo{author}{\bibfnamefont{J.}~\bibnamefont{Alba}},
  \bibinfo{author}{\bibfnamefont{L.}~\bibnamefont{Di~Rienzo}},
  \bibinfo{author}{\bibfnamefont{E.}~\bibnamefont{Milanetti}},
  \bibinfo{author}{\bibfnamefont{O.}~\bibnamefont{Acuto}}, \bibnamefont{and}
  \bibinfo{author}{\bibfnamefont{M.}~\bibnamefont{D’Abramo}},
  \bibinfo{journal}{Cells} \textbf{\bibinfo{volume}{9}}, \bibinfo{pages}{942}
  (\bibinfo{year}{2020}).

\bibitem[{\citenamefont{Han et~al.}(2019)\citenamefont{Han, Sit, Christoffer,
  Chen, and Kihara}}]{han2019global}
\bibinfo{author}{\bibfnamefont{X.}~\bibnamefont{Han}},
  \bibinfo{author}{\bibfnamefont{A.}~\bibnamefont{Sit}},
  \bibinfo{author}{\bibfnamefont{C.}~\bibnamefont{Christoffer}},
  \bibinfo{author}{\bibfnamefont{S.}~\bibnamefont{Chen}}, \bibnamefont{and}
  \bibinfo{author}{\bibfnamefont{D.}~\bibnamefont{Kihara}},
  \bibinfo{journal}{PLoS computational biology} \textbf{\bibinfo{volume}{15}},
  \bibinfo{pages}{e1006969} (\bibinfo{year}{2019}).

\bibitem[{\citenamefont{Di~Rienzo
  et~al.}(2020{\natexlab{b}})\citenamefont{Di~Rienzo, Milanetti, Testi,
  Montemiglio, Baiocco, Boffi, and Ruocco}}]{di2020novel}
\bibinfo{author}{\bibfnamefont{L.}~\bibnamefont{Di~Rienzo}},
  \bibinfo{author}{\bibfnamefont{E.}~\bibnamefont{Milanetti}},
  \bibinfo{author}{\bibfnamefont{C.}~\bibnamefont{Testi}},
  \bibinfo{author}{\bibfnamefont{L.~C.} \bibnamefont{Montemiglio}},
  \bibinfo{author}{\bibfnamefont{P.}~\bibnamefont{Baiocco}},
  \bibinfo{author}{\bibfnamefont{A.}~\bibnamefont{Boffi}}, \bibnamefont{and}
  \bibinfo{author}{\bibfnamefont{G.}~\bibnamefont{Ruocco}},
  \bibinfo{journal}{Computational and structural biotechnology journal}
  \textbf{\bibinfo{volume}{18}}, \bibinfo{pages}{2678}
  (\bibinfo{year}{2020}{\natexlab{b}}).

\bibitem[{\citenamefont{Di~Rienzo
  et~al.}(2021{\natexlab{b}})\citenamefont{Di~Rienzo, Monti, Milanetti, Miotto,
  Boffi, Tartaglia, and Ruocco}}]{di2021computational}
\bibinfo{author}{\bibfnamefont{L.}~\bibnamefont{Di~Rienzo}},
  \bibinfo{author}{\bibfnamefont{M.}~\bibnamefont{Monti}},
  \bibinfo{author}{\bibfnamefont{E.}~\bibnamefont{Milanetti}},
  \bibinfo{author}{\bibfnamefont{M.}~\bibnamefont{Miotto}},
  \bibinfo{author}{\bibfnamefont{A.}~\bibnamefont{Boffi}},
  \bibinfo{author}{\bibfnamefont{G.~G.} \bibnamefont{Tartaglia}},
  \bibnamefont{and} \bibinfo{author}{\bibfnamefont{G.}~\bibnamefont{Ruocco}},
  \bibinfo{journal}{Computational and Structural Biotechnology Journal}
  (\bibinfo{year}{2021}{\natexlab{b}}).

\bibitem[{\citenamefont{Li et~al.}(2003)\citenamefont{Li, Li, Yang, Smith-Gill,
  and Mariuzza}}]{li2003x}
\bibinfo{author}{\bibfnamefont{Y.}~\bibnamefont{Li}},
  \bibinfo{author}{\bibfnamefont{H.}~\bibnamefont{Li}},
  \bibinfo{author}{\bibfnamefont{F.}~\bibnamefont{Yang}},
  \bibinfo{author}{\bibfnamefont{S.~J.} \bibnamefont{Smith-Gill}},
  \bibnamefont{and} \bibinfo{author}{\bibfnamefont{R.~A.}
  \bibnamefont{Mariuzza}}, \bibinfo{journal}{Nature Structural \& Molecular
  Biology} \textbf{\bibinfo{volume}{10}}, \bibinfo{pages}{482}
  (\bibinfo{year}{2003}).

\bibitem[{\citenamefont{Singh et~al.}(2018)\citenamefont{Singh, Tank, Dwiwedi,
  Charan, Kaur, Sidhu, and Chugh}}]{singh2018monoclonal}
\bibinfo{author}{\bibfnamefont{S.}~\bibnamefont{Singh}},
  \bibinfo{author}{\bibfnamefont{N.~K.} \bibnamefont{Tank}},
  \bibinfo{author}{\bibfnamefont{P.}~\bibnamefont{Dwiwedi}},
  \bibinfo{author}{\bibfnamefont{J.}~\bibnamefont{Charan}},
  \bibinfo{author}{\bibfnamefont{R.}~\bibnamefont{Kaur}},
  \bibinfo{author}{\bibfnamefont{P.}~\bibnamefont{Sidhu}}, \bibnamefont{and}
  \bibinfo{author}{\bibfnamefont{V.~K.} \bibnamefont{Chugh}},
  \bibinfo{journal}{Current clinical pharmacology}
  \textbf{\bibinfo{volume}{13}}, \bibinfo{pages}{85} (\bibinfo{year}{2018}).

\bibitem[{\citenamefont{Saeed et~al.}(2017)\citenamefont{Saeed, Wang, Ling, and
  Wang}}]{saeed2017antibody}
\bibinfo{author}{\bibfnamefont{A.~F.} \bibnamefont{Saeed}},
  \bibinfo{author}{\bibfnamefont{R.}~\bibnamefont{Wang}},
  \bibinfo{author}{\bibfnamefont{S.}~\bibnamefont{Ling}}, \bibnamefont{and}
  \bibinfo{author}{\bibfnamefont{S.}~\bibnamefont{Wang}},
  \bibinfo{journal}{Frontiers in microbiology} \textbf{\bibinfo{volume}{8}},
  \bibinfo{pages}{495} (\bibinfo{year}{2017}).

\bibitem[{\citenamefont{Gotwals et~al.}(2017)\citenamefont{Gotwals, Cameron,
  Cipolletta, Cremasco, Crystal, Hewes, Mueller, Quaratino, Sabatos-Peyton,
  Petruzzelli et~al.}}]{gotwals2017prospects}
\bibinfo{author}{\bibfnamefont{P.}~\bibnamefont{Gotwals}},
  \bibinfo{author}{\bibfnamefont{S.}~\bibnamefont{Cameron}},
  \bibinfo{author}{\bibfnamefont{D.}~\bibnamefont{Cipolletta}},
  \bibinfo{author}{\bibfnamefont{V.}~\bibnamefont{Cremasco}},
  \bibinfo{author}{\bibfnamefont{A.}~\bibnamefont{Crystal}},
  \bibinfo{author}{\bibfnamefont{B.}~\bibnamefont{Hewes}},
  \bibinfo{author}{\bibfnamefont{B.}~\bibnamefont{Mueller}},
  \bibinfo{author}{\bibfnamefont{S.}~\bibnamefont{Quaratino}},
  \bibinfo{author}{\bibfnamefont{C.}~\bibnamefont{Sabatos-Peyton}},
  \bibinfo{author}{\bibfnamefont{L.}~\bibnamefont{Petruzzelli}},
  \bibnamefont{et~al.}, \bibinfo{journal}{Nature Reviews Cancer}
  \textbf{\bibinfo{volume}{17}}, \bibinfo{pages}{286} (\bibinfo{year}{2017}).

\bibitem[{\citenamefont{Chothia and Lesk}(1987)}]{chothia1987canonical}
\bibinfo{author}{\bibfnamefont{C.}~\bibnamefont{Chothia}} \bibnamefont{and}
  \bibinfo{author}{\bibfnamefont{A.~M.} \bibnamefont{Lesk}},
  \bibinfo{journal}{Journal of molecular biology}
  \textbf{\bibinfo{volume}{196}}, \bibinfo{pages}{901} (\bibinfo{year}{1987}).

\bibitem[{\citenamefont{Chothia et~al.}(1989)\citenamefont{Chothia, Lesk,
  Tramontano, Levitt, Smith-Gill, Air, Sheriff, Padlan, Davies, Tulip
  et~al.}}]{chothia1989conformations}
\bibinfo{author}{\bibfnamefont{C.}~\bibnamefont{Chothia}},
  \bibinfo{author}{\bibfnamefont{A.~M.} \bibnamefont{Lesk}},
  \bibinfo{author}{\bibfnamefont{A.}~\bibnamefont{Tramontano}},
  \bibinfo{author}{\bibfnamefont{M.}~\bibnamefont{Levitt}},
  \bibinfo{author}{\bibfnamefont{S.~J.} \bibnamefont{Smith-Gill}},
  \bibinfo{author}{\bibfnamefont{G.}~\bibnamefont{Air}},
  \bibinfo{author}{\bibfnamefont{S.}~\bibnamefont{Sheriff}},
  \bibinfo{author}{\bibfnamefont{E.~A.} \bibnamefont{Padlan}},
  \bibinfo{author}{\bibfnamefont{D.}~\bibnamefont{Davies}},
  \bibinfo{author}{\bibfnamefont{W.~R.} \bibnamefont{Tulip}},
  \bibnamefont{et~al.}, \bibinfo{journal}{Nature}
  \textbf{\bibinfo{volume}{342}}, \bibinfo{pages}{877} (\bibinfo{year}{1989}).

\bibitem[{\citenamefont{Tramontano et~al.}(1990)\citenamefont{Tramontano,
  Chothia, and Lesk}}]{tramontano1990framework}
\bibinfo{author}{\bibfnamefont{A.}~\bibnamefont{Tramontano}},
  \bibinfo{author}{\bibfnamefont{C.}~\bibnamefont{Chothia}}, \bibnamefont{and}
  \bibinfo{author}{\bibfnamefont{A.~M.} \bibnamefont{Lesk}},
  \bibinfo{journal}{Journal of molecular biology}
  \textbf{\bibinfo{volume}{215}}, \bibinfo{pages}{175} (\bibinfo{year}{1990}).

\bibitem[{\citenamefont{Chothia et~al.}(1992)\citenamefont{Chothia, Lesk,
  Gherardi, Tomlinson, Walter, Marks, Llewelyn, and
  Winter}}]{chothia1992structural}
\bibinfo{author}{\bibfnamefont{C.}~\bibnamefont{Chothia}},
  \bibinfo{author}{\bibfnamefont{A.~M.} \bibnamefont{Lesk}},
  \bibinfo{author}{\bibfnamefont{E.}~\bibnamefont{Gherardi}},
  \bibinfo{author}{\bibfnamefont{I.~M.} \bibnamefont{Tomlinson}},
  \bibinfo{author}{\bibfnamefont{G.}~\bibnamefont{Walter}},
  \bibinfo{author}{\bibfnamefont{J.~D.} \bibnamefont{Marks}},
  \bibinfo{author}{\bibfnamefont{M.~B.} \bibnamefont{Llewelyn}},
  \bibnamefont{and} \bibinfo{author}{\bibfnamefont{G.}~\bibnamefont{Winter}},
  \bibinfo{journal}{Journal of molecular biology}
  \textbf{\bibinfo{volume}{227}}, \bibinfo{pages}{799} (\bibinfo{year}{1992}).

\bibitem[{\citenamefont{Foote and Winter}(1992)}]{foote1992antibody}
\bibinfo{author}{\bibfnamefont{J.}~\bibnamefont{Foote}} \bibnamefont{and}
  \bibinfo{author}{\bibfnamefont{G.}~\bibnamefont{Winter}},
  \bibinfo{journal}{Journal of molecular biology}
  \textbf{\bibinfo{volume}{224}}, \bibinfo{pages}{487} (\bibinfo{year}{1992}).

\bibitem[{\citenamefont{Decanniere et~al.}(2000)\citenamefont{Decanniere,
  Muyldermans, and Wyns}}]{decanniere2000canonical}
\bibinfo{author}{\bibfnamefont{K.}~\bibnamefont{Decanniere}},
  \bibinfo{author}{\bibfnamefont{S.}~\bibnamefont{Muyldermans}},
  \bibnamefont{and} \bibinfo{author}{\bibfnamefont{L.}~\bibnamefont{Wyns}},
  \bibinfo{journal}{Journal of molecular biology}
  \textbf{\bibinfo{volume}{300}}, \bibinfo{pages}{83} (\bibinfo{year}{2000}).

\bibitem[{\citenamefont{Chailyan et~al.}(2011)\citenamefont{Chailyan,
  Marcatili, Cirillo, and Tramontano}}]{chailyan2011structural}
\bibinfo{author}{\bibfnamefont{A.}~\bibnamefont{Chailyan}},
  \bibinfo{author}{\bibfnamefont{P.}~\bibnamefont{Marcatili}},
  \bibinfo{author}{\bibfnamefont{D.}~\bibnamefont{Cirillo}}, \bibnamefont{and}
  \bibinfo{author}{\bibfnamefont{A.}~\bibnamefont{Tramontano}},
  \bibinfo{journal}{Proteins: Structure, Function, and Bioinformatics}
  \textbf{\bibinfo{volume}{79}}, \bibinfo{pages}{1513} (\bibinfo{year}{2011}).

\bibitem[{\citenamefont{North et~al.}(2011)\citenamefont{North, Lehmann, and
  Dunbrack~Jr}}]{north2011new}
\bibinfo{author}{\bibfnamefont{B.}~\bibnamefont{North}},
  \bibinfo{author}{\bibfnamefont{A.}~\bibnamefont{Lehmann}}, \bibnamefont{and}
  \bibinfo{author}{\bibfnamefont{R.~L.} \bibnamefont{Dunbrack~Jr}},
  \bibinfo{journal}{Journal of molecular biology}
  \textbf{\bibinfo{volume}{406}}, \bibinfo{pages}{228} (\bibinfo{year}{2011}).

\bibitem[{\citenamefont{Dunbar et~al.}(2013)\citenamefont{Dunbar, Krawczyk,
  Leem, Baker, Fuchs, Georges, Shi, and Deane}}]{dunbar2013sabdab}
\bibinfo{author}{\bibfnamefont{J.}~\bibnamefont{Dunbar}},
  \bibinfo{author}{\bibfnamefont{K.}~\bibnamefont{Krawczyk}},
  \bibinfo{author}{\bibfnamefont{J.}~\bibnamefont{Leem}},
  \bibinfo{author}{\bibfnamefont{T.}~\bibnamefont{Baker}},
  \bibinfo{author}{\bibfnamefont{A.}~\bibnamefont{Fuchs}},
  \bibinfo{author}{\bibfnamefont{G.}~\bibnamefont{Georges}},
  \bibinfo{author}{\bibfnamefont{J.}~\bibnamefont{Shi}}, \bibnamefont{and}
  \bibinfo{author}{\bibfnamefont{C.~M.} \bibnamefont{Deane}},
  \bibinfo{journal}{Nucleic acids research} \textbf{\bibinfo{volume}{42}},
  \bibinfo{pages}{D1140} (\bibinfo{year}{2013}).

\bibitem[{\citenamefont{Dunbar et~al.}(2016)\citenamefont{Dunbar, Krawczyk,
  Leem, Marks, Nowak, Regep, Georges, Kelm, Popovic, and
  Deane}}]{dunbar2016sabpred}
\bibinfo{author}{\bibfnamefont{J.}~\bibnamefont{Dunbar}},
  \bibinfo{author}{\bibfnamefont{K.}~\bibnamefont{Krawczyk}},
  \bibinfo{author}{\bibfnamefont{J.}~\bibnamefont{Leem}},
  \bibinfo{author}{\bibfnamefont{C.}~\bibnamefont{Marks}},
  \bibinfo{author}{\bibfnamefont{J.}~\bibnamefont{Nowak}},
  \bibinfo{author}{\bibfnamefont{C.}~\bibnamefont{Regep}},
  \bibinfo{author}{\bibfnamefont{G.}~\bibnamefont{Georges}},
  \bibinfo{author}{\bibfnamefont{S.}~\bibnamefont{Kelm}},
  \bibinfo{author}{\bibfnamefont{B.}~\bibnamefont{Popovic}}, \bibnamefont{and}
  \bibinfo{author}{\bibfnamefont{C.~M.} \bibnamefont{Deane}},
  \bibinfo{journal}{Nucleic acids research} \textbf{\bibinfo{volume}{44}},
  \bibinfo{pages}{W474} (\bibinfo{year}{2016}).

\bibitem[{\citenamefont{Lepore et~al.}(2017)\citenamefont{Lepore, Olimpieri,
  Messih, and Tramontano}}]{lepore2017pigspro}
\bibinfo{author}{\bibfnamefont{R.}~\bibnamefont{Lepore}},
  \bibinfo{author}{\bibfnamefont{P.~P.} \bibnamefont{Olimpieri}},
  \bibinfo{author}{\bibfnamefont{M.~A.} \bibnamefont{Messih}},
  \bibnamefont{and}
  \bibinfo{author}{\bibfnamefont{A.}~\bibnamefont{Tramontano}},
  \bibinfo{journal}{Nucleic Acids Research} \textbf{\bibinfo{volume}{45}},
  \bibinfo{pages}{W17} (\bibinfo{year}{2017}).

\bibitem[{\citenamefont{Weitzner et~al.}(2017)\citenamefont{Weitzner,
  Jeliazkov, Lyskov, Marze, Kuroda, Frick, Adolf-Bryfogle, Biswas, Dunbrack,
  and Gray}}]{weitzner2017modeling}
\bibinfo{author}{\bibfnamefont{B.~D.} \bibnamefont{Weitzner}},
  \bibinfo{author}{\bibfnamefont{J.~R.} \bibnamefont{Jeliazkov}},
  \bibinfo{author}{\bibfnamefont{S.}~\bibnamefont{Lyskov}},
  \bibinfo{author}{\bibfnamefont{N.}~\bibnamefont{Marze}},
  \bibinfo{author}{\bibfnamefont{D.}~\bibnamefont{Kuroda}},
  \bibinfo{author}{\bibfnamefont{R.}~\bibnamefont{Frick}},
  \bibinfo{author}{\bibfnamefont{J.}~\bibnamefont{Adolf-Bryfogle}},
  \bibinfo{author}{\bibfnamefont{N.}~\bibnamefont{Biswas}},
  \bibinfo{author}{\bibfnamefont{R.~L.} \bibnamefont{Dunbrack}},
  \bibnamefont{and} \bibinfo{author}{\bibfnamefont{J.~J.} \bibnamefont{Gray}},
  \bibinfo{journal}{Nature protocols} \textbf{\bibinfo{volume}{12}},
  \bibinfo{pages}{401} (\bibinfo{year}{2017}).

\bibitem[{\citenamefont{Abanades et~al.}(2022)\citenamefont{Abanades, Georges,
  Bujotzek, and Deane}}]{Abanades2022ABlooper}
\bibinfo{author}{\bibfnamefont{B.}~\bibnamefont{Abanades}},
  \bibinfo{author}{\bibfnamefont{G.}~\bibnamefont{Georges}},
  \bibinfo{author}{\bibfnamefont{A.}~\bibnamefont{Bujotzek}}, \bibnamefont{and}
  \bibinfo{author}{\bibfnamefont{C.~M.} \bibnamefont{Deane}},
  \bibinfo{journal}{Bioinformatics}  (\bibinfo{year}{2022}), ISSN
  \bibinfo{issn}{1367-4803}, \bibinfo{note}{btac016},
  \eprint{https://academic.oup.com/bioinformatics/advance-article-pdf/doi/10.1093/bioinformatics/btac016/42377836/btac016.pdf},
  \urlprefix\url{https://doi.org/10.1093/bioinformatics/btac016}.

\bibitem[{\citenamefont{Olimpieri et~al.}(2013)\citenamefont{Olimpieri,
  Chailyan, Tramontano, and Marcatili}}]{olimpieri2013prediction}
\bibinfo{author}{\bibfnamefont{P.~P.} \bibnamefont{Olimpieri}},
  \bibinfo{author}{\bibfnamefont{A.}~\bibnamefont{Chailyan}},
  \bibinfo{author}{\bibfnamefont{A.}~\bibnamefont{Tramontano}},
  \bibnamefont{and}
  \bibinfo{author}{\bibfnamefont{P.}~\bibnamefont{Marcatili}},
  \bibinfo{journal}{Bioinformatics} \textbf{\bibinfo{volume}{29}},
  \bibinfo{pages}{2285} (\bibinfo{year}{2013}).

\bibitem[{\citenamefont{Liberis et~al.}(2018)\citenamefont{Liberis,
  Veli{\v{c}}kovi{\'c}, Sormanni, Vendruscolo, and
  Li{\`o}}}]{liberis2018parapred}
\bibinfo{author}{\bibfnamefont{E.}~\bibnamefont{Liberis}},
  \bibinfo{author}{\bibfnamefont{P.}~\bibnamefont{Veli{\v{c}}kovi{\'c}}},
  \bibinfo{author}{\bibfnamefont{P.}~\bibnamefont{Sormanni}},
  \bibinfo{author}{\bibfnamefont{M.}~\bibnamefont{Vendruscolo}},
  \bibnamefont{and} \bibinfo{author}{\bibfnamefont{P.}~\bibnamefont{Li{\`o}}},
  \bibinfo{journal}{Bioinformatics} \textbf{\bibinfo{volume}{34}},
  \bibinfo{pages}{2944} (\bibinfo{year}{2018}).

\bibitem[{\citenamefont{Potocnakova et~al.}(2016)\citenamefont{Potocnakova,
  Bhide, and Pulzova}}]{potocnakova2016introduction}
\bibinfo{author}{\bibfnamefont{L.}~\bibnamefont{Potocnakova}},
  \bibinfo{author}{\bibfnamefont{M.}~\bibnamefont{Bhide}}, \bibnamefont{and}
  \bibinfo{author}{\bibfnamefont{L.~B.} \bibnamefont{Pulzova}},
  \bibinfo{journal}{Journal of immunology research}
  \textbf{\bibinfo{volume}{2016}} (\bibinfo{year}{2016}).

\bibitem[{\citenamefont{Norman et~al.}(2020)\citenamefont{Norman, Ambrosetti,
  Bonvin, Colwell, Kelm, Kumar, and Krawczyk}}]{norman2020computational}
\bibinfo{author}{\bibfnamefont{R.~A.} \bibnamefont{Norman}},
  \bibinfo{author}{\bibfnamefont{F.}~\bibnamefont{Ambrosetti}},
  \bibinfo{author}{\bibfnamefont{A.~M.} \bibnamefont{Bonvin}},
  \bibinfo{author}{\bibfnamefont{L.~J.} \bibnamefont{Colwell}},
  \bibinfo{author}{\bibfnamefont{S.}~\bibnamefont{Kelm}},
  \bibinfo{author}{\bibfnamefont{S.}~\bibnamefont{Kumar}}, \bibnamefont{and}
  \bibinfo{author}{\bibfnamefont{K.}~\bibnamefont{Krawczyk}},
  \bibinfo{journal}{Briefings in bioinformatics} \textbf{\bibinfo{volume}{21}},
  \bibinfo{pages}{1549} (\bibinfo{year}{2020}).

\bibitem[{\citenamefont{Pantazes and Maranas}(2010)}]{pantazes2010optcdr}
\bibinfo{author}{\bibfnamefont{R.}~\bibnamefont{Pantazes}} \bibnamefont{and}
  \bibinfo{author}{\bibfnamefont{C.~D.} \bibnamefont{Maranas}},
  \bibinfo{journal}{Protein Engineering, Design \& Selection}
  \textbf{\bibinfo{volume}{23}}, \bibinfo{pages}{849} (\bibinfo{year}{2010}).

\bibitem[{\citenamefont{Li et~al.}(2014)\citenamefont{Li, Pantazes, and
  Maranas}}]{li2014optmaven}
\bibinfo{author}{\bibfnamefont{T.}~\bibnamefont{Li}},
  \bibinfo{author}{\bibfnamefont{R.~J.} \bibnamefont{Pantazes}},
  \bibnamefont{and} \bibinfo{author}{\bibfnamefont{C.~D.}
  \bibnamefont{Maranas}}, \bibinfo{journal}{PloS one}
  \textbf{\bibinfo{volume}{9}}, \bibinfo{pages}{e105954}
  (\bibinfo{year}{2014}).

\bibitem[{\citenamefont{Adolf-Bryfogle
  et~al.}(2018)\citenamefont{Adolf-Bryfogle, Kalyuzhniy, Kubitz, Weitzner, Hu,
  Adachi, Schief, and Dunbrack~Jr}}]{adolf2018rosettaantibodydesign}
\bibinfo{author}{\bibfnamefont{J.}~\bibnamefont{Adolf-Bryfogle}},
  \bibinfo{author}{\bibfnamefont{O.}~\bibnamefont{Kalyuzhniy}},
  \bibinfo{author}{\bibfnamefont{M.}~\bibnamefont{Kubitz}},
  \bibinfo{author}{\bibfnamefont{B.~D.} \bibnamefont{Weitzner}},
  \bibinfo{author}{\bibfnamefont{X.}~\bibnamefont{Hu}},
  \bibinfo{author}{\bibfnamefont{Y.}~\bibnamefont{Adachi}},
  \bibinfo{author}{\bibfnamefont{W.~R.} \bibnamefont{Schief}},
  \bibnamefont{and} \bibinfo{author}{\bibfnamefont{R.~L.}
  \bibnamefont{Dunbrack~Jr}}, \bibinfo{journal}{PLoS computational biology}
  \textbf{\bibinfo{volume}{14}}, \bibinfo{pages}{e1006112}
  (\bibinfo{year}{2018}).

\bibitem[{\citenamefont{Lapidoth et~al.}(2015)\citenamefont{Lapidoth, Baran,
  Pszolla, Norn, Alon, Tyka, and Fleishman}}]{lapidoth2015abdesign}
\bibinfo{author}{\bibfnamefont{G.~D.} \bibnamefont{Lapidoth}},
  \bibinfo{author}{\bibfnamefont{D.}~\bibnamefont{Baran}},
  \bibinfo{author}{\bibfnamefont{G.~M.} \bibnamefont{Pszolla}},
  \bibinfo{author}{\bibfnamefont{C.}~\bibnamefont{Norn}},
  \bibinfo{author}{\bibfnamefont{A.}~\bibnamefont{Alon}},
  \bibinfo{author}{\bibfnamefont{M.~D.} \bibnamefont{Tyka}}, \bibnamefont{and}
  \bibinfo{author}{\bibfnamefont{S.~J.} \bibnamefont{Fleishman}},
  \bibinfo{journal}{Proteins: Structure, Function, and Bioinformatics}
  \textbf{\bibinfo{volume}{83}}, \bibinfo{pages}{1385} (\bibinfo{year}{2015}).

\bibitem[{\citenamefont{Milanetti
  et~al.}(2021{\natexlab{b}})\citenamefont{Milanetti, Miotto, Rienzo, Nagaraj,
  Monti, Golbek, Gosti, Roeters, Weidner, Otzen et~al.}}]{milanetti2020silico}
\bibinfo{author}{\bibfnamefont{E.}~\bibnamefont{Milanetti}},
  \bibinfo{author}{\bibfnamefont{M.}~\bibnamefont{Miotto}},
  \bibinfo{author}{\bibfnamefont{L.~D.} \bibnamefont{Rienzo}},
  \bibinfo{author}{\bibfnamefont{M.}~\bibnamefont{Nagaraj}},
  \bibinfo{author}{\bibfnamefont{M.}~\bibnamefont{Monti}},
  \bibinfo{author}{\bibfnamefont{T.~W.} \bibnamefont{Golbek}},
  \bibinfo{author}{\bibfnamefont{G.}~\bibnamefont{Gosti}},
  \bibinfo{author}{\bibfnamefont{S.~J.} \bibnamefont{Roeters}},
  \bibinfo{author}{\bibfnamefont{T.}~\bibnamefont{Weidner}},
  \bibinfo{author}{\bibfnamefont{D.~E.} \bibnamefont{Otzen}},
  \bibnamefont{et~al.}, \bibinfo{journal}{Frontiers in Molecular Biosciences}
  \textbf{\bibinfo{volume}{8}} (\bibinfo{year}{2021}{\natexlab{b}}).

\bibitem[{\citenamefont{Miotto et~al.}(2021)\citenamefont{Miotto, Rienzo,
  B{\`{o}}, Boffi, Ruocco, and Milanetti}}]{Miotto2021Molecular}
\bibinfo{author}{\bibfnamefont{M.}~\bibnamefont{Miotto}},
  \bibinfo{author}{\bibfnamefont{L.~D.} \bibnamefont{Rienzo}},
  \bibinfo{author}{\bibfnamefont{L.}~\bibnamefont{B{\`{o}}}},
  \bibinfo{author}{\bibfnamefont{A.}~\bibnamefont{Boffi}},
  \bibinfo{author}{\bibfnamefont{G.}~\bibnamefont{Ruocco}}, \bibnamefont{and}
  \bibinfo{author}{\bibfnamefont{E.}~\bibnamefont{Milanetti}},
  \bibinfo{journal}{Frontiers in Molecular Biosciences}
  \textbf{\bibinfo{volume}{8}} (\bibinfo{year}{2021}).

\bibitem[{\citenamefont{B{\`o} et~al.}(2021)\citenamefont{B{\`o}, Miotto,
  Di~Rienzo, Milanetti, and Ruocco}}]{bo2021exploring}
\bibinfo{author}{\bibfnamefont{L.}~\bibnamefont{B{\`o}}},
  \bibinfo{author}{\bibfnamefont{M.}~\bibnamefont{Miotto}},
  \bibinfo{author}{\bibfnamefont{L.}~\bibnamefont{Di~Rienzo}},
  \bibinfo{author}{\bibfnamefont{E.}~\bibnamefont{Milanetti}},
  \bibnamefont{and} \bibinfo{author}{\bibfnamefont{G.}~\bibnamefont{Ruocco}},
  \bibinfo{journal}{Frontiers in medical technology} p.~\bibinfo{pages}{24}
  (\bibinfo{year}{2021}).

\bibitem[{\citenamefont{Miotto et~al.}(2022)\citenamefont{Miotto, Di~Rienzo,
  Gosti, Bo, Parisi, Piacentini, Boffi, Ruocco, and
  Milanetti}}]{miotto2022inferring}
\bibinfo{author}{\bibfnamefont{M.}~\bibnamefont{Miotto}},
  \bibinfo{author}{\bibfnamefont{L.}~\bibnamefont{Di~Rienzo}},
  \bibinfo{author}{\bibfnamefont{G.}~\bibnamefont{Gosti}},
  \bibinfo{author}{\bibfnamefont{L.}~\bibnamefont{Bo}},
  \bibinfo{author}{\bibfnamefont{G.}~\bibnamefont{Parisi}},
  \bibinfo{author}{\bibfnamefont{R.}~\bibnamefont{Piacentini}},
  \bibinfo{author}{\bibfnamefont{A.}~\bibnamefont{Boffi}},
  \bibinfo{author}{\bibfnamefont{G.}~\bibnamefont{Ruocco}}, \bibnamefont{and}
  \bibinfo{author}{\bibfnamefont{E.}~\bibnamefont{Milanetti}},
  \bibinfo{journal}{Communications Biology} \textbf{\bibinfo{volume}{5}},
  \bibinfo{pages}{1} (\bibinfo{year}{2022}).

\bibitem[{\citenamefont{Richards}(1977)}]{richards1977areas}
\bibinfo{author}{\bibfnamefont{F.~M.} \bibnamefont{Richards}},
  \bibinfo{journal}{Annual review of biophysics and bioengineering}
  \textbf{\bibinfo{volume}{6}}, \bibinfo{pages}{151} (\bibinfo{year}{1977}).

\bibitem[{\citenamefont{Kirkpatrick et~al.}(1983)\citenamefont{Kirkpatrick,
  Gelatt, and Vecchi}}]{kirkpatrick1983optimization}
\bibinfo{author}{\bibfnamefont{S.}~\bibnamefont{Kirkpatrick}},
  \bibinfo{author}{\bibfnamefont{C.~D.} \bibnamefont{Gelatt}},
  \bibnamefont{and} \bibinfo{author}{\bibfnamefont{M.~P.}
  \bibnamefont{Vecchi}}, \bibinfo{journal}{science}
  \textbf{\bibinfo{volume}{220}}, \bibinfo{pages}{671} (\bibinfo{year}{1983}).

\bibitem[{\citenamefont{Huang et~al.}(2020)\citenamefont{Huang, Wang, Li, Ren,
  Zhao, Hu, Zhang, Fan, Xu, Gu et~al.}}]{huang2020clinical}
\bibinfo{author}{\bibfnamefont{C.}~\bibnamefont{Huang}},
  \bibinfo{author}{\bibfnamefont{Y.}~\bibnamefont{Wang}},
  \bibinfo{author}{\bibfnamefont{X.}~\bibnamefont{Li}},
  \bibinfo{author}{\bibfnamefont{L.}~\bibnamefont{Ren}},
  \bibinfo{author}{\bibfnamefont{J.}~\bibnamefont{Zhao}},
  \bibinfo{author}{\bibfnamefont{Y.}~\bibnamefont{Hu}},
  \bibinfo{author}{\bibfnamefont{L.}~\bibnamefont{Zhang}},
  \bibinfo{author}{\bibfnamefont{G.}~\bibnamefont{Fan}},
  \bibinfo{author}{\bibfnamefont{J.}~\bibnamefont{Xu}},
  \bibinfo{author}{\bibfnamefont{X.}~\bibnamefont{Gu}}, \bibnamefont{et~al.},
  \bibinfo{journal}{The Lancet} \textbf{\bibinfo{volume}{395}},
  \bibinfo{pages}{497} (\bibinfo{year}{2020}).

\bibitem[{\citenamefont{Zhu et~al.}(2020)\citenamefont{Zhu, Zhang, Wang, Li,
  Yang, Song, Zhao, Huang, Shi, Lu et~al.}}]{zhu2020novel}
\bibinfo{author}{\bibfnamefont{N.}~\bibnamefont{Zhu}},
  \bibinfo{author}{\bibfnamefont{D.}~\bibnamefont{Zhang}},
  \bibinfo{author}{\bibfnamefont{W.}~\bibnamefont{Wang}},
  \bibinfo{author}{\bibfnamefont{X.}~\bibnamefont{Li}},
  \bibinfo{author}{\bibfnamefont{B.}~\bibnamefont{Yang}},
  \bibinfo{author}{\bibfnamefont{J.}~\bibnamefont{Song}},
  \bibinfo{author}{\bibfnamefont{X.}~\bibnamefont{Zhao}},
  \bibinfo{author}{\bibfnamefont{B.}~\bibnamefont{Huang}},
  \bibinfo{author}{\bibfnamefont{W.}~\bibnamefont{Shi}},
  \bibinfo{author}{\bibfnamefont{R.}~\bibnamefont{Lu}}, \bibnamefont{et~al.},
  \bibinfo{journal}{New England Journal of Medicine}  (\bibinfo{year}{2020}).

\bibitem[{\citenamefont{Zhou et~al.}(2020)\citenamefont{Zhou, Yang, Wang, Hu,
  Zhang, Zhang, Si, Zhu, Li, Huang et~al.}}]{zhou2020pneumonia}
\bibinfo{author}{\bibfnamefont{P.}~\bibnamefont{Zhou}},
  \bibinfo{author}{\bibfnamefont{X.-L.} \bibnamefont{Yang}},
  \bibinfo{author}{\bibfnamefont{X.-G.} \bibnamefont{Wang}},
  \bibinfo{author}{\bibfnamefont{B.}~\bibnamefont{Hu}},
  \bibinfo{author}{\bibfnamefont{L.}~\bibnamefont{Zhang}},
  \bibinfo{author}{\bibfnamefont{W.}~\bibnamefont{Zhang}},
  \bibinfo{author}{\bibfnamefont{H.-R.} \bibnamefont{Si}},
  \bibinfo{author}{\bibfnamefont{Y.}~\bibnamefont{Zhu}},
  \bibinfo{author}{\bibfnamefont{B.}~\bibnamefont{Li}},
  \bibinfo{author}{\bibfnamefont{C.-L.} \bibnamefont{Huang}},
  \bibnamefont{et~al.}, \bibinfo{journal}{Nature} pp. \bibinfo{pages}{1--4}
  (\bibinfo{year}{2020}).

\bibitem[{\citenamefont{Walls et~al.}(2020)\citenamefont{Walls, Park,
  Tortorici, Wall, McGuire, and Veesler}}]{walls2020structure}
\bibinfo{author}{\bibfnamefont{A.~C.} \bibnamefont{Walls}},
  \bibinfo{author}{\bibfnamefont{Y.-J.} \bibnamefont{Park}},
  \bibinfo{author}{\bibfnamefont{M.~A.} \bibnamefont{Tortorici}},
  \bibinfo{author}{\bibfnamefont{A.}~\bibnamefont{Wall}},
  \bibinfo{author}{\bibfnamefont{A.~T.} \bibnamefont{McGuire}},
  \bibnamefont{and} \bibinfo{author}{\bibfnamefont{D.}~\bibnamefont{Veesler}},
  \bibinfo{journal}{Cell}  (\bibinfo{year}{2020}).

\bibitem[{\citenamefont{Wan et~al.}(2020)\citenamefont{Wan, Shang, Graham,
  Baric, and Li}}]{wan2020receptor}
\bibinfo{author}{\bibfnamefont{Y.}~\bibnamefont{Wan}},
  \bibinfo{author}{\bibfnamefont{J.}~\bibnamefont{Shang}},
  \bibinfo{author}{\bibfnamefont{R.}~\bibnamefont{Graham}},
  \bibinfo{author}{\bibfnamefont{R.~S.} \bibnamefont{Baric}}, \bibnamefont{and}
  \bibinfo{author}{\bibfnamefont{F.}~\bibnamefont{Li}},
  \bibinfo{journal}{Journal of virology}  (\bibinfo{year}{2020}).

\bibitem[{\citenamefont{Raybould et~al.}(2021)\citenamefont{Raybould,
  Kovaltsuk, Marks, and Deane}}]{raybould2021cov}
\bibinfo{author}{\bibfnamefont{M.~I.} \bibnamefont{Raybould}},
  \bibinfo{author}{\bibfnamefont{A.}~\bibnamefont{Kovaltsuk}},
  \bibinfo{author}{\bibfnamefont{C.}~\bibnamefont{Marks}}, \bibnamefont{and}
  \bibinfo{author}{\bibfnamefont{C.~M.} \bibnamefont{Deane}},
  \bibinfo{journal}{Bioinformatics} \textbf{\bibinfo{volume}{37}},
  \bibinfo{pages}{734} (\bibinfo{year}{2021}).

\bibitem[{\citenamefont{Li et~al.}(2017)\citenamefont{Li, Hulswit, Widjaja,
  Raj, McBride, Peng, Widagdo, Tortorici, Van~Dieren, Lang
  et~al.}}]{li2017identification}
\bibinfo{author}{\bibfnamefont{W.}~\bibnamefont{Li}},
  \bibinfo{author}{\bibfnamefont{R.~J.} \bibnamefont{Hulswit}},
  \bibinfo{author}{\bibfnamefont{I.}~\bibnamefont{Widjaja}},
  \bibinfo{author}{\bibfnamefont{V.~S.} \bibnamefont{Raj}},
  \bibinfo{author}{\bibfnamefont{R.}~\bibnamefont{McBride}},
  \bibinfo{author}{\bibfnamefont{W.}~\bibnamefont{Peng}},
  \bibinfo{author}{\bibfnamefont{W.}~\bibnamefont{Widagdo}},
  \bibinfo{author}{\bibfnamefont{M.~A.} \bibnamefont{Tortorici}},
  \bibinfo{author}{\bibfnamefont{B.}~\bibnamefont{Van~Dieren}},
  \bibinfo{author}{\bibfnamefont{Y.}~\bibnamefont{Lang}}, \bibnamefont{et~al.},
  \bibinfo{journal}{Proceedings of the National Academy of Sciences}
  \textbf{\bibinfo{volume}{114}}, \bibinfo{pages}{E8508}
  (\bibinfo{year}{2017}).

\bibitem[{\citenamefont{Baker et~al.}(2020)\citenamefont{Baker, Richards, Guy,
  Congdon, Hasan, Zwetsloot, Gallo, Lewandowski, Stansfeld, Straube
  et~al.}}]{baker2020sars}
\bibinfo{author}{\bibfnamefont{A.~N.} \bibnamefont{Baker}},
  \bibinfo{author}{\bibfnamefont{S.-J.} \bibnamefont{Richards}},
  \bibinfo{author}{\bibfnamefont{C.~S.} \bibnamefont{Guy}},
  \bibinfo{author}{\bibfnamefont{T.~R.} \bibnamefont{Congdon}},
  \bibinfo{author}{\bibfnamefont{M.}~\bibnamefont{Hasan}},
  \bibinfo{author}{\bibfnamefont{A.~J.} \bibnamefont{Zwetsloot}},
  \bibinfo{author}{\bibfnamefont{A.}~\bibnamefont{Gallo}},
  \bibinfo{author}{\bibfnamefont{J.~R.} \bibnamefont{Lewandowski}},
  \bibinfo{author}{\bibfnamefont{P.~J.} \bibnamefont{Stansfeld}},
  \bibinfo{author}{\bibfnamefont{A.}~\bibnamefont{Straube}},
  \bibnamefont{et~al.}, \bibinfo{journal}{ACS central science}
  \textbf{\bibinfo{volume}{6}}, \bibinfo{pages}{2046} (\bibinfo{year}{2020}).

\bibitem[{\citenamefont{Huang et~al.}(2010)\citenamefont{Huang, Niu, Gao, Fu,
  and Li}}]{huang2010cd}
\bibinfo{author}{\bibfnamefont{Y.}~\bibnamefont{Huang}},
  \bibinfo{author}{\bibfnamefont{B.}~\bibnamefont{Niu}},
  \bibinfo{author}{\bibfnamefont{Y.}~\bibnamefont{Gao}},
  \bibinfo{author}{\bibfnamefont{L.}~\bibnamefont{Fu}}, \bibnamefont{and}
  \bibinfo{author}{\bibfnamefont{W.}~\bibnamefont{Li}},
  \bibinfo{journal}{Bioinformatics} \textbf{\bibinfo{volume}{26}},
  \bibinfo{pages}{680} (\bibinfo{year}{2010}).

\bibitem[{\citenamefont{Krivov et~al.}(2009)\citenamefont{Krivov, Shapovalov,
  and Dunbrack~Jr}}]{krivov2009improved}
\bibinfo{author}{\bibfnamefont{G.~G.} \bibnamefont{Krivov}},
  \bibinfo{author}{\bibfnamefont{M.~V.} \bibnamefont{Shapovalov}},
  \bibnamefont{and} \bibinfo{author}{\bibfnamefont{R.~L.}
  \bibnamefont{Dunbrack~Jr}}, \bibinfo{journal}{Proteins: Structure, Function,
  and Bioinformatics} \textbf{\bibinfo{volume}{77}}, \bibinfo{pages}{778}
  (\bibinfo{year}{2009}).

\bibitem[{\citenamefont{MacKerell~Jr et~al.}(2002)\citenamefont{MacKerell~Jr,
  Brooks, Brooks~III, Nilsson, Roux, Won, and Karplus}}]{mackerell2002charmm}
\bibinfo{author}{\bibfnamefont{A.~D.} \bibnamefont{MacKerell~Jr}},
  \bibinfo{author}{\bibfnamefont{B.}~\bibnamefont{Brooks}},
  \bibinfo{author}{\bibfnamefont{C.~L.} \bibnamefont{Brooks~III}},
  \bibinfo{author}{\bibfnamefont{L.}~\bibnamefont{Nilsson}},
  \bibinfo{author}{\bibfnamefont{B.}~\bibnamefont{Roux}},
  \bibinfo{author}{\bibfnamefont{Y.}~\bibnamefont{Won}}, \bibnamefont{and}
  \bibinfo{author}{\bibfnamefont{M.}~\bibnamefont{Karplus}},
  \bibinfo{journal}{Encyclopedia of computational chemistry}
  \textbf{\bibinfo{volume}{1}} (\bibinfo{year}{2002}).


\end{thebibliography}

\end{document}